\documentclass[prb,twocolumn,superscriptaddress,floatfix,amsmath,amssymb]{revtex4}
\usepackage{multirow}
\usepackage{graphicx}
\usepackage{bm}
\usepackage{dcolumn}
\usepackage{picture}
\usepackage[hidelinks]{hyperref}
\usepackage{natbib}
\usepackage{amsmath}
\usepackage{times}
\usepackage{color}
\usepackage{xcolor}
\usepackage{soul}
\usepackage{amsmath}
\usepackage{amssymb}

\newcommand{\la}{La$_4$Ni$_3$O$_{10}$}
\newcommand{\pr}{Pr$_4$Ni$_3$O$_{10}$}
\newcommand{\nd}{Nd$_4$Ni$_3$O$_{10}$}
\newcommand{\re}{$R_4$Ni$_3$O$_{10}$}

\newcommand{\tcdw}{$T_{\rm MMT}$}

\newcommand{\dcp}{$\Delta c_{\rm p}$}

\newcommand{\jmk}{J/(mol\,K)}

\newcommand{\alb}{$\alpha^{\rm bgr}$}

\newcommand{\etal}{\textit{et~al.}}

\begin{document}
	\title{Structural and physical properties of trilayer nickelates $R_4$Ni$_3$O$_{10}$ ($R =$ La, Pr and Nd)}

	\author{Dibyata Rout}
	\affiliation{Department of Physics, Indian Institute of Science Education and Research, Pune, Maharashtra-411008, India}
	
	\author{Sanchayeta Ranajit Mudi}
	\affiliation{Department of Physics, Indian Institute of Science Education and Research, Pune, Maharashtra-411008, India}
	
	\author{Marco Hoffmann}
	\affiliation{Kirchhoff Institute of Physics, Heidelberg University, INF 227, D-69120, Heidelberg, Germany}
	
	\author{Sven Spachmann}
	\affiliation{Kirchhoff Institute of Physics, Heidelberg University, INF 227, D-69120, Heidelberg, Germany}
	
	\author{R\"{u}diger Klingeler}
	\affiliation{Kirchhoff Institute of Physics, Heidelberg University, INF 227, D-69120, Heidelberg, Germany}\affiliation{ Centre for Advanced  Materials (CAM), Heidelberg University, INF 227, D-69120, Heidelberg, Germany}
	
	\author{Surjeet Singh}
	\email[email:]{surjeet.singh@iiserpune.ac.in}
	\affiliation{Department of Physics, Indian Institute of Science Education and Research, Pune, Maharashtra-411008, India}\affiliation{Center for Energy Science, Indian Institute of Science Education and Research, Pune, Maharashtra-411008, India}
	
	\date{\today}
\begin{abstract}
			{
	Here, we investigate in detail the low temperature structural and physical properties of the trilayer nickelates \re~($R =$ La, Pr and Nd). We show that all three nickelates crystallize with a monoclinic symmetry (space group $P2_1\slash a, Z =4$), and undergo a metal-to-metal  transition (MMT) near \tcdw~= $135$~K (La), $156$~K (Pr) and $160$~K (Nd). The lattice parameters show a distinct anomaly at \tcdw in all cases, however, without any signs of lattice symmetry breaking. Unambiguous signatures of MMT are observed in magnetic susceptibility and transport, suggesting a strong coupling between the electronic, magnetic and structural degrees of freedom. Analysis of thermal expansion yields hydrostatic pressure dependence of MMT in close agreement with experimental results. In \pr, there is a possible antiferromagnetic ordering of the Pr$^{3+}$ ions in the rocksalt (RS) layers near $5$ K, much suppressed compared to $\theta_p \sim-36$ K. In contrast, Pr$^{3+}$ ions in the perovskite-block (PB) layers exhibit a non-magnetic singlet ground state. In \nd, the CF ground state of Nd$^{3+}$ ions in both RS and PB layers is a Kramers doublet. The heat capacity of \nd~shows a pronounced Schottky-like anomaly near $T = 40$ K; a sharp upturn indicating short-range correlations between the Nd moments is also observed below $10$ K. However, no signs of long-range ordering could be found down to $2$ K despite a sizeable value of $\theta_p \sim-40$ K. The strongly suppressed magnetic long-range ordering in both $R = $ Pr and Nd suggests the presence of strong magnetic frustration in these compounds. The low-temperature resistivity in all cases shows a $-\sqrt{T}$ dependence. Due to the presence of an overwhelming Schottky contribution, the electronic term in the specific heat of \pr\ amd \nd~appears to be hugely inflated, which can be \textit{falsely} interpreted as a sign of heavy fermion behavior as was done in a recent study on \nd.	
				
			}
	\end{abstract}
		
\maketitle

 \section{Introduction}
 \label{Introduction}
The transition metal oxides based on nickel, or the nickelates for short, have witnessed a resurgence of interest in the last few years. Several recent papers have shown that nickelates are unique due to their strongly coupled charge, spin and lattice degrees of freedom, which can be manipulated to engineer novel electronic and magnetic phases (see for example Ref.~\citenum{PhysRevLett.103.156401, Zhao2014, PhysRevLett.113.227206}). Another reason for this resurgence can be attributed to the discovery of superconductivity in Nd$_{0.8}$Sr$_{0.2}$NiO$_{2}$ by Li et al. in the year 2019, which, in fact, led to the fulfillment of a long-sought-after quest for superconductivity in the nickelates \cite{Li2019}. Nearly two years before this momentous discovery, an ARPES study on single crystals of La$_4$Ni$_3$O$_{10}$, which is the $n =3$ member of the Ruddlesden Popper (RP) La$_{n+1}$Ni$_n$O$_{3n+1}$ $(n = 1, 2, 3 \dots \infty)$ series, revealed a large hole Fermi surface that closely resembles the Fermi surface of optimally hole-doped cuprates~\cite{Li2017} (see also Ref. \citenum{Zhang2017}). This discovery is important since the infinite layer NdNiO$_{2}$ (called the $T'$ phase) is related to the perovskite NdNiO$_3$ ($n = \infty$ member of RP series) from which it is obtained through a process of chemical reduction. In general, there is a whole range of infinite layer $T'$ phases given by $R_{n+1}$Ni$_n$O$_{2n+2}$ $(n = 1, 2, 3 \dots \infty)$, where $R$ is usually an alkaline earth or rare-earth ion, that are analogously related to their corresponding RP $R_{n+1}$Ni$_n$O$_{3n+1}$ phases. The nickelates of the RP series, therefore, constitute the primary phases with perovskite-type structure elements from which other nickelates, including the infinite layer $T'$ variants, can be derived.

A survey of past literature on the nickelates of the RP series reveals that the $n = 1, 2, 3. \dots$ members of the $RP$ series are relatively much less investigated -- an exception to this being La$_2$NiO$_{4-\delta}$ ($n = 1$), which shows an interesting phase diagram as a function of $\delta$ (see for example: Ref. \citenum{GREENBLATT1997174}). These intermediate members between $n = 1$ and $n = \infty$, in fact, exhibit a mixed-valent state, ranging from $2+$ for $ n = 1$ to $3+$ for $n = \infty$. Such a mixed-valency is well-known to give rise to strongly coupled electronic and magnetic phases (see for example Ref. \citenum{CoeyAIP1999}). Hence, there is a significant interest to study them in the recent years. 

In particular, the $n = 3$ member of the RP series, consisting of the compounds \re\ ($R =$ La, Pr and Nd) with an average Ni valence of $2.67$, will be interesting to investigate. The compounds $R_4$Ni$_3$O$_{10}$ ($R =$ La, Pr and Nd) are relatively easy to synthesize in pure form, and can also be readily reduced to their corresponding infinite layer $T'$ analogs. Previous studies have shown that they undergo a metal-to-metal transition (MMT) in the temperature range $135$ K to $160$ K depending on the identity of the $R$ ion. Recently, they have drawn a considerable attention (see for example: Ref. \citenum{Li2017,zhang2020intertwined, PhysRevB.97.115116, PhysRevB.101.195142, PhysRevB.101.104104, ZhangPRM2020, HuangfuPRR2020}). However,  majority of these studies have mainly focused on understanding the nature of MMT where Ni 3d electrons play a crucial role. The magnetic ground state of the rare-earth sublattice or of the $4f$ electrons, and the interplay of $3d$ and $4f$ electrons have not been studied in detail so far. Moreover, the question of whether there is a structural phase transition associated with MMT or not has remain unsettled issues over the years.

Here, we investigate the resistivity, thermopower, thermal conductivity, magnetic susceptibility and specific heat of $R_4$Ni$_3$O$_{10}$ ($R = $La, Pr and Nd) in considerable detail to explore and understand the low-temperature properties arising due to 4f electrons. Further, to throw light on the nature of MMT, the crystal structure of $R_4$Ni$_3$O$_{10}$ ($R =$ La, Pr and Nd) is examined over a broad temperature range spanning MMT in all three compounds using a very high-resolution synchrotron data on high-quality samples. This is complemented by high-resolution capacitance dilatometry to investigate the temperature dependence of thermal expansion and Gr\"{u}neisen parameter across the MMT.
	
We show that in Pr$_4$Ni$_3$O$_{10}$, the Pr-moments in the rock-salt block layers undergo a magnetic ordering near $T_N = 5$ K while the Pr$^{3+}$ ions in the perovskite block layers exhibit a crystal field split non-magnetic singlet ground state. On the other hand, Nd$^{3+}$ moments in Nd$_4$Ni$_3$O$_{10}$ show no long-range ordering down to $2$ K (lowest temperature in our measurements). The paramagnetic Curie-temperatures for these compounds is found to lie in the range $-40$ K to $-50$ K indicating the presence of strong magnetic frustration. The effective carrier mass deduced from specific heat and thermopower lies in the range 2 to 4 times the free electron mass indicating moderately enhanced electronic correlations. The resistivities of all three compounds show an upturn at low-temperatures obeying a $-\sqrt{T}$ dependence. No evidence for the Kondo effect or the heavy Fermion behavior in any of the $R_4$Ni$_3$O$_{10}$'s could be found, contradicting the claim of heavy fermion state due to Ni$^{3+}$ centered Kondo effect in \nd\ published recently \cite{PhysRevB.101.195142}.

The rest of the paper has been organized as follows: The details of experimental methods are given in section \ref{ED}. This is followed by Results and Discussion section (\ref{III}), which has been divided further into several subsections for convenience. The details of crystal structure appear under \ref{IIIA}. The electrical and thermal transport, and  the magnetic susceptibility have been briefly discussed in \ref{ET}. This is followed by subsections on specific heat (\ref{SH}) and thermal expansion (\ref{TE}) both of which form the crux of the paper. Finally, a summary of the important results, and conclusions drawn are presented under section \ref{SC}. 

\section{Experimental Details}
 \label{ED}
 Conventional solid state synthesis of the higher members of the RP family leads to the formation of mixed phases and intergrowth~\cite{A702424J,ZHANG1994402}, which greatly influences the physical properties of the compounds. Hence, we adopted a wet chemical method to synthesize these compounds in pure phase. Further details of sample preparation are given here (see Ref.~\citenum{SM}). The phase purity was monitored using a Bruker D8 Advance powder X-ray diffractometer. The chemical composition of the samples was analyzed using the energy dispersive X-ray analysis (EDX) technique in a Zeiss Ultra Plus scanning electron microscope. Since the structural and electronic properties of RP phases often show strong dependence on the oxygen stoichiometry, we carried out complete decomposition of our samples under $10\%$ Ar-H$_2$ atmosphere employing a heating rate of $5$ K/min in a high resolution TGA setup (Netzsch STA $449$ F1). Using these experiments, we inferred the oxygen stoichiometry to lie in the range $97~\%$ to $98~\%$ of the ideal value for all the samples.

The high-resolution synchrotron powder X-ray diffraction experiments were carried out at the MSPD-BLO4 beamline of the ALBA synchrotron center, Barcelona, Spain. The samples were prepared in the form of finely ground powders that were placed in a borosilicate capillary tube of $0.5$ mm inner diameter. The sample was cooled using an Oxford Cryostream $700$ series nitrogen blower, and the diffractograms were collected in the range $0^o\leq 2\theta\leq30^o$ with a step size of $0.003^o$. The incident beam energy was set at $38$ keV ($\lambda =0.3263$~\AA) and a high resolution MAD$26$ detector with an angular  resolution of about $4 \cdot 10^{-4}$ was used to resolve any subtle structural modifications~\cite{Fauth2015}. The data at each temperature was collected at a rate of $30$ min/scan. The structural refinement was done by the Rietveld method using the FULLPROF suite~\cite{rodriguez1993recent}. During the refinement, the occupancies of the O--sites are fixed as fully occupied as X-Ray diffraction is not sensitive to the position of lighter elements. The Pseudo-Voigt function was used to model the line-profile. Linear interpolation method was used to define the background. To account for the anisotropic strain broadening of the peaks, the Broadening Model (quartic form) was used. In this model, only certain $h k l$ dependent strain parameters ($S_{hkl}$) were refined corresponding to the Laue class used. Further details ae given in the SI. The quality of the refinement was assessed, both from the visual inspection of the fitted pattern or the difference plots, and the quantitative assessment on the basis of $\chi^2$, and the R-factors ($R_\mathrm{WP}, R_\mathrm{EXP}$ and $R_\mathrm{P}$). For fitting the low temperature data, the lattice parameters were refined along with angle $\beta$, the overall isotropic displacement factor ($B_\mathrm{iso}$), and the strain coefficients.

  Magnetization, resistivity, thermopower and specific heat measurements were done using a Physical Property Measurement System (PPMS), Quantum Design USA. Magnetization measurements were done, both, under the zero-field-cooled (ZFC) and field-cooled (FC) conditions. Resistivity measurements were done on sintered rectangular samples of known dimensions using the four-probe method. Gold wires were used for electrical contacts with silver conducting paste. Specific-heat measurements were performed using the relaxation method in the PPMS. The heat capacity of the sample holder and APIEZON N grease (addenda) was determined prior to the measurements.

 The relative length changes $dL/L$ were studied on cuboid shaped sintered samples of approximate dimensions $3 \times 2 \times 1~$mm$^{3}$. The measurements were done in zero magnetic field by means of a three-terminal high-resolution capacitance dilatometer\cite{kuchler2012compact}. The relative volume changes $dV/V= 3 dL/L$ and the volume thermal expansion coefficient $\beta=3\alpha$, with $\alpha = 1/L\cdot dL(T)/dT$ are derived.

 \begin{figure}[!]
 	\centering
 	\includegraphics[width= 1.1\columnwidth]{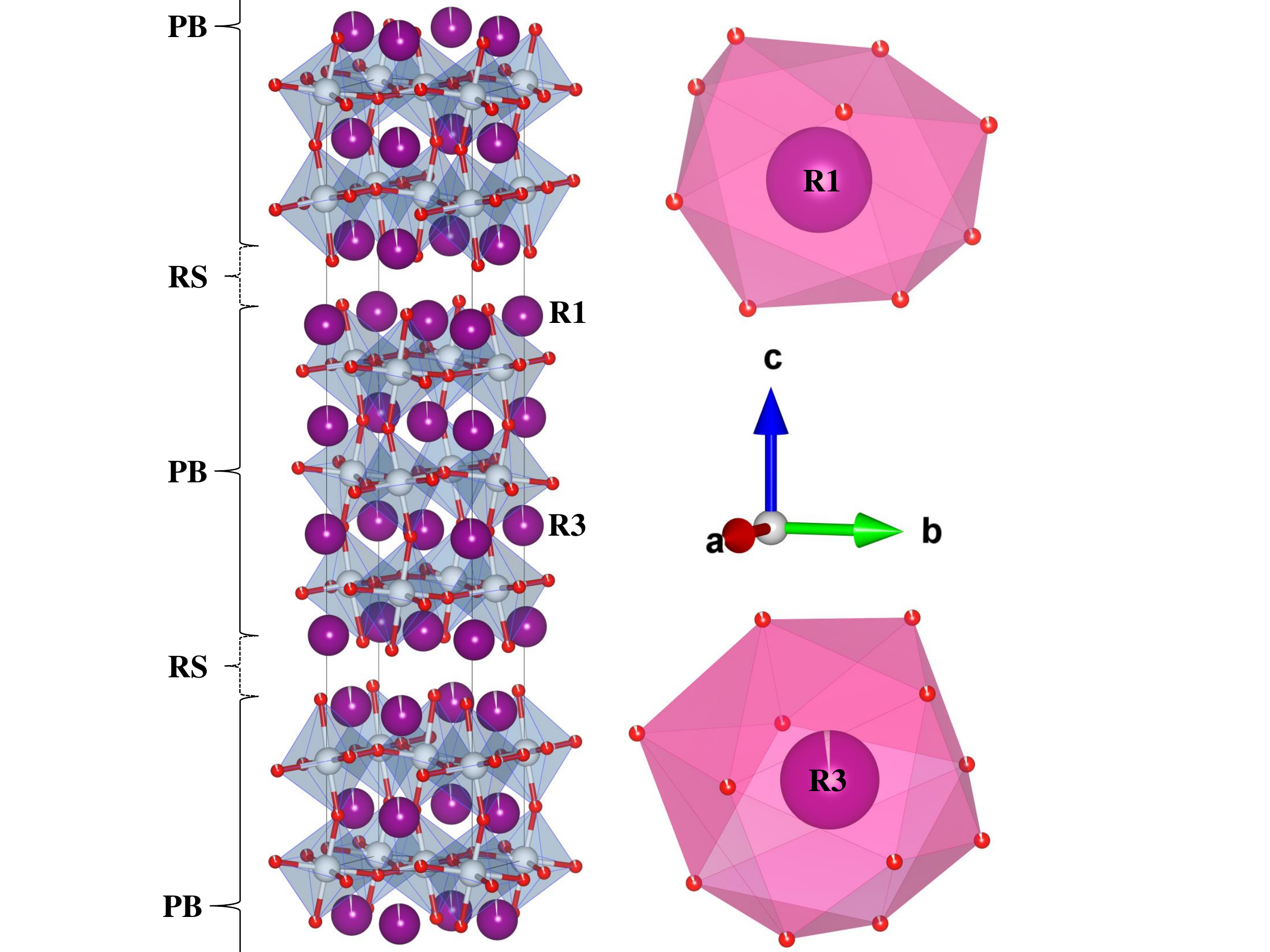}
 	\caption {The crystal structure of trilayer $R_4$Ni$_3$O$_{10}$ ($R =$ La, Pr and Nd) nickelates. Here PB represents the perovskite block and RS represents the rocksalt layer. R1 and R3 denote $9-$fold and $12-$fold coordinated rare-earth ions located in RS and PB, respectively.}
 	\label{CS}
 \end{figure}

\section{Results and Discussion}
 \label{III}
\subsection{Crystal structure}
\label{IIIA}
\textit{La$_4$Ni$_3$O$_{10}$}: There is a great deal of ambiguity in previous literature regarding the space group that correctly defines the crystal structure of La$_4$Ni$_3$O$_{10}$. The earliest work by Sepp\"{a}nen et al. reported an orthorhombic space group $Fmmm$ \cite{seppanen1979crystal}. However, Tkalich et al. \cite{tkalich1991synthesis}, and Voronin et al. \cite{VORONIN2001202} used the space group $Cmca$. Ling et al. \cite{LING2000517}, on the other hand, found the orthorhombic space group $Bmab$ (unconventional setting for $Cmca$) to be more suitable for refining their neutron powder diffraction data. Zhang et al. carried out structural refinement on the powders obtained by crushing high-pressure floating-zone grown single crystalline specimens \cite{ZhangPRM2020}. They propose that La$_4$Ni$_3$O$_{10}$ crystallizes in a mixture of $Bmab$ and $P2_1\slash a$---the phase fraction between the two phases being a function of the cooling condition employed \cite{zhang2019high}. For example, the phase $Bmab$ transforms almost completely to $P2_1\slash a$ when annealed under flowing oxygen. Finally, in a recent synchrotron based study by Kumar et al., the space group symmetry $P2_1/a, Z = 4$ has been endorsed~\cite{KUMAR2020165915}.

    \begin{figure*}[!]
	\centering
	\includegraphics[width= 2\columnwidth]{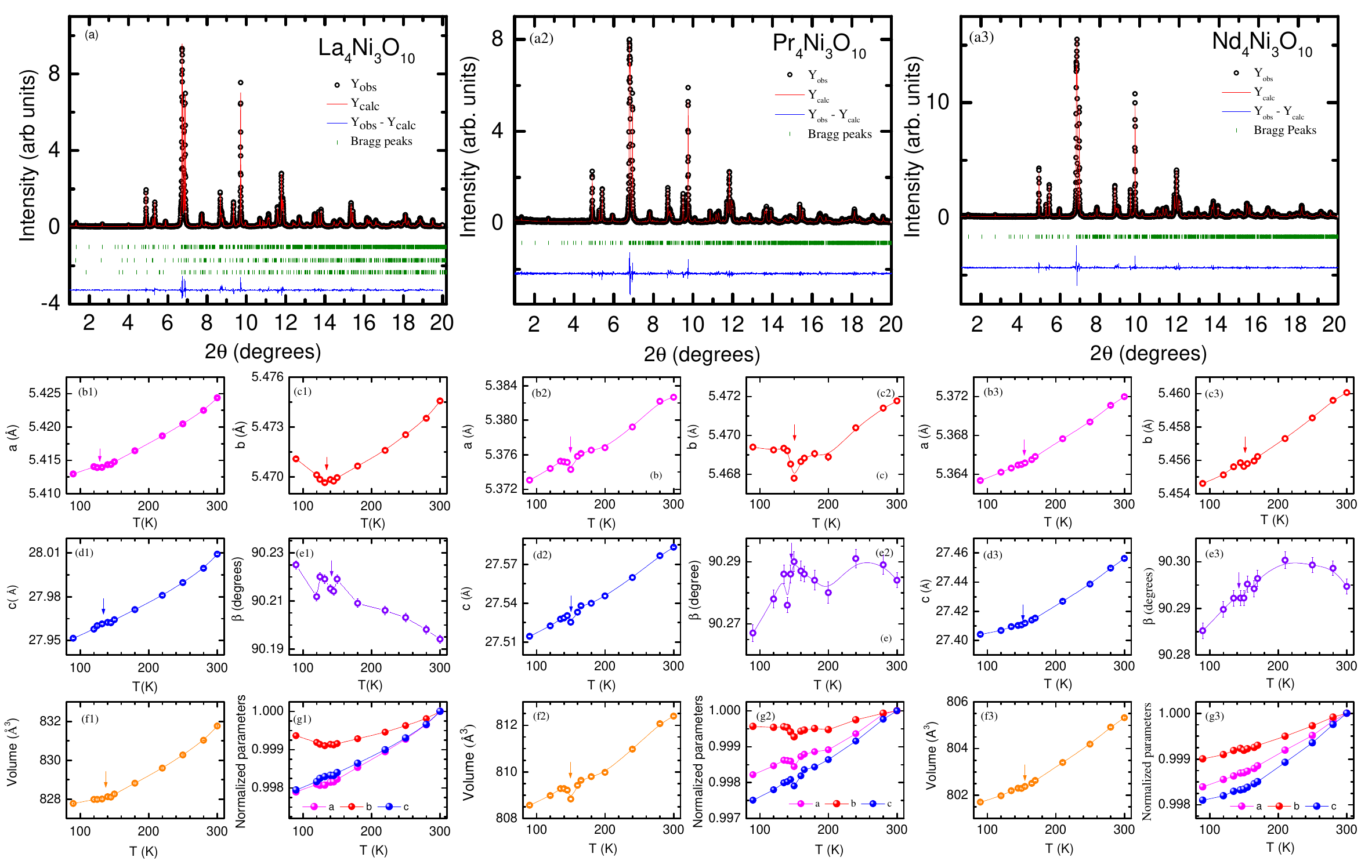}
	\caption {Rietveld refinement results of the room temperature synchrotron powder X-ray diffraction data for the three nickelates: (a1) La$_4$Ni$_3$O$_{10}$, (a2)~\pr\, (a3)~\nd. The black circles represent observed data; the red lines is the calculated intensity, and the vertical green bars indicate the positions of the Bragg peaks; the blue line at the bottom is the difference plot. In panel a1 the first, second, and third row of Bragg peaks correspond to $P2_1\slash a$, $Bmab$ and La$_3$Ni$_2$O$_7$ phases, respectively. In panels a2 and a3, only single phase refinement is done and the Bragg peaks (vertical green bars) are for $P2_1\slash a$. Panels b's, c's \& d's show the temperature variation of lattice parameters $a$, $b$ and $c$, respectively; panels e's \& f's show the temperature dependence of angle $\beta$ and unit cell volume, respectively; panels (g's) shows the normalized unit cell parameters. In some cases the size of the error bars is smaller than  that of the data points.}
	\label{RR4310}
\end{figure*}

In order to find the most appropriate space group from among those that were previously reported, we started by refining the structure using one space group at a time. To avoid biasing this procedure, every space group is tried till the refinement could not be improved further. Using this procedure (see supplementary Material for details \cite{SM}), we found that $P2_1\slash a$ (SG no.~$14, Z = 4$) best fits the experimental data. However, even with $P2_1\slash a$,  the calculated profile around the high intensity peaks in the range $2\theta = 6 ^\circ$ to $7^\circ$, and those around $2\theta = 9.7 ^\circ$, remains far from perfect as shown in the SI. In the paper by Kumar et al.~\cite{KUMAR2020165915} also a similar difference between the calculated and measured intensities can be seen (see Fig. $2a$ and $2b$ of Ref. \citenum{KUMAR2020165915}).

\begin{figure}
 	\centering
 	\includegraphics[width=0.47\textwidth]{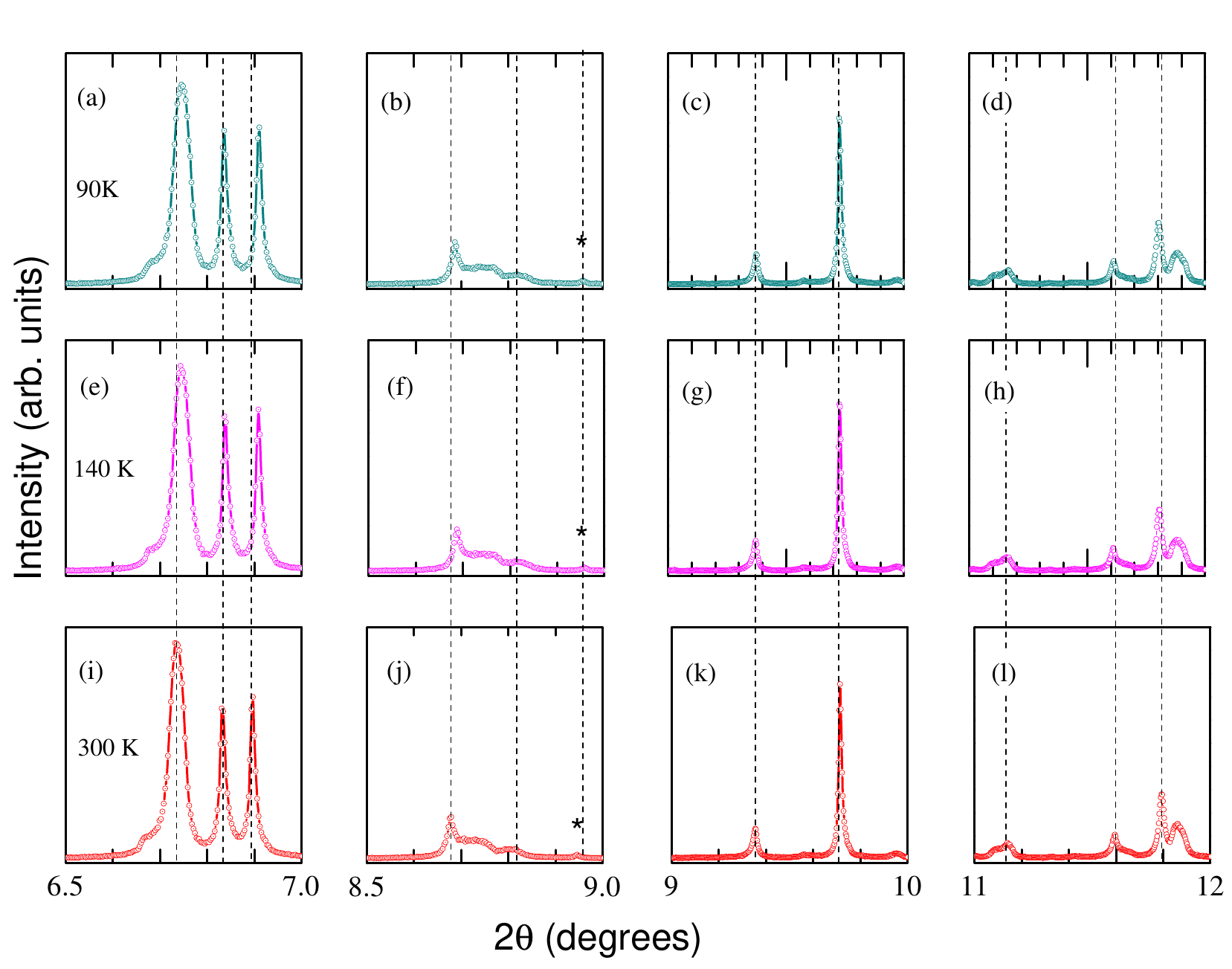}
 	\caption {Synchrotron powder X-ray diffraction data for La$_4$Ni$_3$O$_{10}$ at three representative temperatures: $90$ K (top row), $140$ K (middle row) and $300$ K (bottom row) over select $2\theta$ ranges. The dashed vertical lines are shown as a guide to the eye. The y-scale in each panel is kept the same. Asterisk indicates an unidentified peak. Analogous low-temperature data for \pr~and \nd~are shown in the Supplementary Material \cite{SM}.}
 	\label{LTXRDLa}
 \end{figure}

 We therefore attempted a mixed phase refinement wherein, besides the principal $P2_1\slash a$ phase, two additional phases: (i) the orthorhombic $Bmab$ (SG no. $64$) phase, and (ii) a lower ($n = 2$) member La$_3$Ni$_2$O$_7$, with an orthorhombic space group $Cmcm$ (SG no. $63$), are also incorporated. As shown in Supplementary Material \cite{SM}, inclusion of the $Bmab$  phase alone improves the quality of fit significantly with $P2_1\slash a : Bmab \equiv 86.3 : 13.7$. In order to see if we can get an even better match with the observed intensities, La$_3$Ni$_2$O$_7$ was also incorporated which lead to a further slight improvement. In this case, we find the ratio of three phases to be $P2_1\slash a : Bmab :$~La$_3$Ni$_2$O$_7 \equiv 85.6: 7.8 : 6.6$. Clearly, in both $2$-- and $3$--phase refinements, the phase fraction of the primary phase $P2_1\slash a$ remains more or less unchanged. Since the $R-$factors quantifying the quality of fit are slightly lower for the $3$--phase refinement, here we have shown the results for the same in Fig.~\ref{RR4310}(a1). Finally, even in the $3-$phase refinement some mismatch between the observed and calculated intensities around  $2\theta = 10 ^\circ$ remains; this has been reported in the previous studies also and may arise from stacking faults \cite{NAGELL20177}. It should also be remarked, that a small extra peak, $\sim 1\%$ of the intensity of the main peak, near $2\theta = 8.95^\circ$, is also observed (see, Fig. \ref{LTXRDLa}(b), \ref{LTXRDLa}(f) or \ref{LTXRDLa}(j)), which indicates the presence of a small unidentified parasitic phase. 
 
 Fig.~\ref{RR4310}(b1-f1) show the temperature variation of the lattice parameters of the $P2_1\slash a$ phase. The lattice parameters decrease monotonically upon cooling exhibiting clearly discernible anomalies at \tcdw. The $b-$axis, in fact, undergoes an expansion upon further cooling below \tcdw. The diffraction patterns recorded below \tcdw\ reveal neither the appearance of any new diffraction peak nor any peak splitting, which suggests that the structural reorganization across the MMT, if any, is rather subtle without any noticeable change of the lattice symmetry (see, Fig. \ref{LTXRDLa}). The negative thermal expansion along the $b-$axis is in agreement with that reported by Kumar et al. (Ref. \citenum{KUMAR2020165915}). The temperature variation of angle $\beta$, shown in panel \ref{RR4310}(e1), shows an increasing behavior upon cooling with a perceptible dip at \tcdw. For comparison, the normalized lattice parameters are shown in Fig.~\ref{RR4310}(g1).

 \textit{Pr$_4$Ni$_3$O$_{10}$}: Fig. \ref{RR4310}(a2) shows the results of Rietveld refinement for Pr$_4$Ni$_3$O$_{10}$. In this case, the refinement was done using the monoclinic space group $P2_1\slash a$ (SG no. $14, Z = 4$) alone, which resulted in a satisfactory fit except near the highest intensity peak where the calculated profile does not exactly match the observed data. Inclusion of strain improved the fitting to some extent but did not resolve the issue completely. Similar inconsistency over the same $2\theta$ range has also been previously observed~\cite{zhang2019high}. Whether the stacking faults or the intergrowth of lower RP members is the reason could not however be reliably ascertained. Also, analogous to La$_4$Ni$_3$O$_{10}$, some intensity mismatch is observed near $2\theta = 10^\circ$ (peak $\bar{2} 2 1$), which may be due to the stacking faults~\cite{NAGELL20177}. 

As shown in Fig.~\ref{RR4310}(b2-e2), in the temperature range around $156$ K, where MMT is expected to occur, a clear anomaly in the lattice parameters is observed. The $b-$axis parameter shows an increase upon cooling below the MMT, analogous to La$_4$Ni$_3$O$_{10}$. In the temperature dependence of angle $\beta$, an appreciable non-monotonic variation has also been observed between the MMT and room-temperature.

 \textit{Nd$_4$Ni$_3$O$_{10}$}: Fig.~\ref{RR4310}(a3) shows the results of Rietveld refinement for Nd$_4$Ni$_3$O$_{10}$ at room-temperature. The structural refinement in this case too is done using the monoclinic space group $P2_1\slash a$ (SG no. $14$; $Z = 4$) alone. Though all the observed peaks could be satisfactorily accounted for, the highest intensity peak was found to be unusually broad and the strain model $2$ is used to account for it (see section \ref{ED}). 

  As shown in Fig.~\ref{RR4310}(b3-e3) the lattice parameters of \nd\ decrease monotonically upon cooling with a weak anomaly around $160$~K, which coincides with \tcdw\ previously reported for this compound. This anomaly is most prominent in the variation of the $b-$parameter. However, unlike the case of \la\ and \pr\, the $b-$parameter in this case continues to decrease upon cooling below the MMT. The temperature variation of angle $\beta$ is shown in panel~\ref{RR4310}(e3). Upon cooling below room temperature, $\beta$ first increases down to about $T = 200$~K and decreases upon further cooling showing a broad peak near $T = 200$~K which may indicate the presence of a rather continuous but subtle and non-monotonic structure evolution occurring even above the MMT, analogous to the case of \pr. However, this should be further confirmed by collecting data at intermediate temperatures for all the samples. 

 \begin{table*}[!]
 	\setlength{\tabcolsep}{4pt}
 	\caption{Refinement parameters obtained using the high-resolution synchrotron data for room the temperature crystal structure of $R_4$Ni$_3$O$_{10}$ ($R = La, Pr$ and $Nd$). The error bar in the lattice parameters is estimated to be of the order of $\pm 0.0002$ in the fourth decimal place}
 	\label{RD}
 	\begin{center}
 		\resizebox{\textwidth}{!}{%
 			\begin{tabular}{c c c c c c c c c c c c c c}
 				\hline\hline
 				Specimen & Space group & SG No. & Phase Type & Phase \% & a($\AA$) & b($\AA)$ & c($\AA$) &$\beta$& $\chi^2$ & $R\rm_{WP}$ & $R\rm_{EXP}$ & $R\rm_P$\\
 				\hline
 				\tabularnewline
 				$La_4Ni_3O_{10}$ & $P2_1\slash a$ & 14 & M$^\dagger$ & 85.6 & 5.4243(5) & 5.4748(5) & 28.0053(4) & $90.192^o(3)$ & 6.15 & 14.7 & 5.90 & 11.7\\
 				& $Bmab$ & 64 & O$^\dagger$ & 7.8 & 5.4040 & 5.4621 & 28.5542 & $90^o$ &  &  &  &\\
 				& $Cmcm$ & 63 & O$^\dagger$ & 6.6 & 20.1250 & 5.4638 & 5.4638 & $90^o$ &  &  &  &\\
 				\tabularnewline
 				\hline
 				\tabularnewline
 				$Pr_4Ni_3O_{10}$ & $P2_1\slash a$ & 14 & M$^\dagger$ & 100  & 5.3826(4) & 5.4717(4) & 27.583(4) & $90.284^o(3)$ & 3.86 & 19.0 & 9.67 & 16\\
 				\tabularnewline
 				\hline
 				\tabularnewline
 				$Nd_4Ni_3O_{10}$ & $P2_1\slash a$ & 14 & M$^\dagger$ & 100 & 5.3719(4) & 5.46(5) & 27.4560(4) & $90.299^o(3)$ & 4.57 & 15.8 & 7.41 & 12.7\\
 				\tabularnewline
 				\hline\hline
 				
 			\end{tabular}
 		   }
 	\end{center}
 \footnotetext{M$^\dagger$ : monoclinic and O$^\dagger$ : orthorhombic}
 \end{table*}

  Table~\ref{RD} summarizes the refinement details for the room temperature crystal structures of $R_4$Ni$_3$O$_{10}$, $R = $ La, Pr and Nd. The room temperature lattice parameters for all three samples are listed in Table \ref{RD}, which agree well with the values reported in previous literature \cite{KUMAR2020165915, BASSAT1998173, OLAFSEN200046}. The crystal structure of  $R_4$Ni$_3$O$_{10}$ (monoclinic $P2_1/a$, $Z = 4$ ), shown in Fig.~\ref{CS}, comprises triple perovskite block (PB) layers ($R$NiO$_3$)$_3$, which consist of corner-linked NiO$_6$ octahedra. These triple PB layers are separated by $R$O layers with the rocksalt (RS) structure. There are four inequivalent $R-$atoms, two of these are located within the PB layers ($R3$, $R4$). They have a deformed $12-$fold coordination analogous to the perovskites $R$NiO$_3$ as shown in Fig. \ref{CS}. The remaining two $R-$atoms are located within the RS layers ($R1$, $R2$) with a $9-$fold coordination. Likewise, there are four distinct crystallographic sites for the Ni atoms. Borrowing the terminology used in Ref. \citenum{OLAFSEN200046}, we shall label them as Ni$1$, Ni$2$: located in the inner layer (ILs), and Ni$3$, Ni$4$: located in the outer layer (OL) that faces the $R$O layer on one side and PB layer on the other. The various R--O and Ni--O bond distances for all three samples are given in the Supplementary Material \cite{SM}. In all three cases, the elongated Ni--O bonds are apical, pointing towards the RS layer which is speculated to be a consequence of Ni$^{3+}$(OL)-Ni$^{2+}$(IL) charge ordering \cite{OLAFSEN200046}.\\

 \subsection {Transport and Magnetization}
\label{ET}
\begin{figure}[!]
  	\centering
  	\includegraphics[width = 0.95\columnwidth]{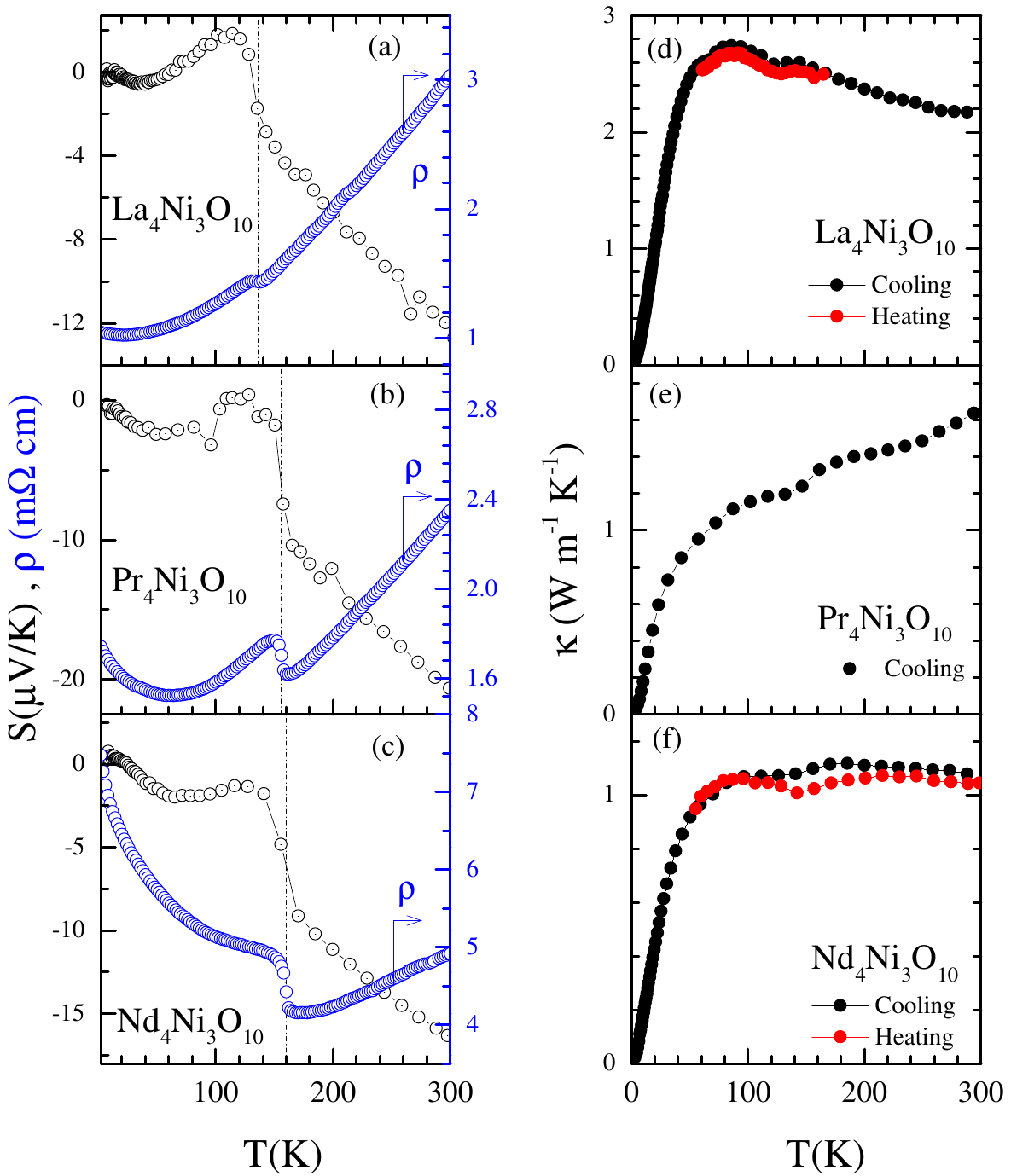}
  	\caption{Panels (a), (b) and (c) show the temperature variation of resistivity ($\rho$) and thermoelectric power ($S$) for \la~, \pr~and \nd~respectively. The temperature variation of their thermal conductivity ($\kappa$) is shown, respectively, in (d), (e) and (f).}
  	\label{Transport}
  \end{figure}

  \textit{Resistivity}: Fig.~\ref{Transport} shows the temperature dependence of resistivity ($\rho$), thermopower ($S$), and thermal conductivity ($\kappa$) for all three samples. We first examine the electrical resistivity. Upon cooling below room temperature, $\rho(T)$ for all three samples decreases monotonically down to a temperature of approximately $136$~K (La), $156$~K (Pr) and $160$~K (Nd). Upon further cooling, $\rho$ increases in a step like fashion, which can be identified with the MMT. The temperature at which the step occurs (\tcdw), agrees well with the temperature where the lattice parameters show an anomaly. The resistivity discontinuity ($\Delta \rho$) at \tcdw~ appears to be first-order like, however, no measurable thermal hysteresis at \tcdw\ could be observed between the heating and cooling data. 

   Below the MMT, the resistivity for \la\ and \pr\ continue to decrease down to some temperature $T_0$, which is followed by an upturn or a region of negative $d\rho/dT$ that persists down to $2$ K.  $T_0$ is $\simeq20$ K and $\simeq80$ K for La$_4$Ni$_3$O$_{10}$ and Pr$_4$Ni$_3$O$_{10}$, respectively. These observations concerning behavior of $\rho(T)$ in the La and Pr compounds are in good agreement with previous reports \cite{SAKURAI201327, BASSAT1998173, KUMAR2020165915, PhysRevB.101.195142}. 
   In Nd$_4$Ni$_3$O$_{10}$, however, $d\rho/dT \simeq 0$ down to about $100$ K, and $<0$ upon further cooling followed by a steep increase below about $50$ K. The upturn in this case is also more pronounced than for La and Pr. In previous resistivity data for \nd\ however a region of negative $d\rho/dT$ for $T <$ \tcdw $< 50$ K is shown~\cite{PhysRevB.101.195142, ZHANG1995236}. Such variations may however arise if there are slight differences in the oxygen off-stoichiometry between the various samples, assuming that other factors, such as, purity and density are same~\cite{BASSAT1998173}. Typically, the extent of oxygen off-stoichiometry is controlled by the synthesis protocol. The \nd~sample used in Ref. ~\citenum{PhysRevB.101.195142} was reported to have been prepared under a pressurized oxygen atmosphere of $5$ bar at $1100^\circ C$ for $24$ h. Similarly, in Ref.~\citenum{ZHANG1995236}, the sample was prepared by annealing it under an oxygen flow for a period of close to $120$ h as opposed to $24$ h in our case. From the TGA data (Fig. 1 of the Supplementary Material \cite{SM}), it is clear that our Nd sample is oxygen deficient with O$_{9.8}$ as its oxygen content instead of full O$_{10}$. The exact oxygen off-stoichiometry for the samples used in Refs.  ~\citenum{PhysRevB.101.195142} and ~\citenum{ZHANG1995236} is not known.  

\begin{figure}[!]
  	\centering
  	\includegraphics[width = \columnwidth]{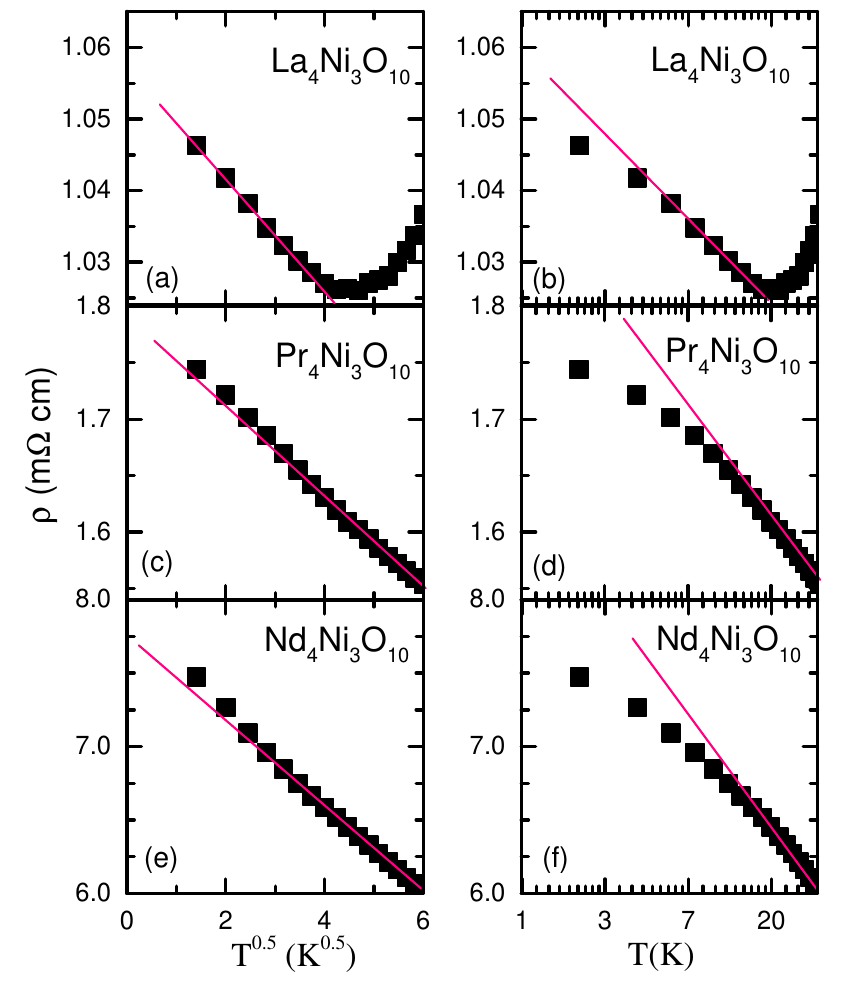}
  	\caption{Temperature ($T$) variation of resistivity ($\rho$) for \la~(a and b), \pr~(c and d) and \nd~(e and f) is plotted on a $T^{0.5}$ scale (left panels), and a $\ln$T scale (right panels) to show that $T^{0.5}$ is a better fit to the data except at low-temperatures ($T < 10$ K) which can be attributed to the short-range ordering of the Pr or Nd moments (see text for details).}
  	\label{Res_fit}
  \end{figure}

   In previous studies, the low-temperature upturn has been variously interpreted. While it is attributed to the weak localization due to inelastic electron-electron interactions in Ref. \citenum{KUMAR2020165915}, the Kondo effect was claimed to be the reason in Ref.~\citenum{PhysRevB.101.195142}. In order to resolve this issue, we replotted the low-temperature data for all three compounds on two different temperature scales: (i) $T^{0.5}$, and (ii) $\ln T$ scales. The results are shown in Fig.~\ref{Res_fit}. Clearly, the data for all three samples are best described by a $-\sqrt{T}$ dependence which persists down to the lowest temperature of $2$~K. Very slight departure from this scaling for \pr~ and \nd~near $10$~K can be attributed to the short-range ordering of the rare-earth moments (\textit{vide infra}). On the contrary, the $-\ln T$ behavior does not describe the upturn in $\rho$ satisfactorily or does so only over a narrow temperature range, with significant departure at low temperatures. Attempts to fit the low-temperature upturn to the Arrhenius or Variable Range Hopping (VRH) models (with or without interactions) also did not give satisfactory results (not shown). The analysis above clearly favors a $-\sqrt{T}$ dependence over other functional dependences commonly used to describe the low-temperature upturn in resistivity. The validity of  $-\sqrt{T}$ behavior suggests that at low-temperatures weak localization due to inelastic electron-electron scattering is possibly what causes the resistivity upturn in all three compounds, which is typical of disordered metals and alloys \cite{RevModPhys.57.287}. Here, the structural disorder might be in the form of stacking faults and intergrowth whose presence is reflected in the powder X-ray diffraction. This conclusion is also in agreement with Ref. \citenum{KUMAR2020165915}. On the other hand, the evidence for the Kondo effect in our data is rather weak.

 \textit{Thermopower}: The thermopower of these samples is shown in panel (a), (b) and (c) of Fig.~\ref{Transport}. The overall behavior and the range of variation of $S$ for the three samples is comparable to that previously reported \cite{SREEDHAR1994208,BASSAT1998173}. The temperature variation of $S$ parallels that of $\rho$ in the sense that at \tcdw , $S(T)$ exhibits a sharp jump, which can be understood based on the Mott's formula for thermopower, which is given by:

\begin{equation}
\centering
S = \frac{\pi ^2}{3} \frac{k_B ^2T}{e} \left ( \frac{\partial \ln\sigma(E)}{\partial E} \right) _{E ={E_F}}
\label{eq1}
\end{equation}

 where $k_B$ is the Boltzmann constant, $\sigma (E)$ is the electrical conductivity, $e$ the electronic charge, and $E_F$ is the Fermi energy. Since, $\sigma$ can be expressed as: $\sigma = n(E)q\mu (E)$, where $n(E) = D(E)f(E)$: $D(E)$ is the density of states, and $f(E)$ the Fermi-Dirac distribution function, and $\mu$ is the carrier mobility, one can rewrite eq. \ref{eq1} with a term in $S$ proportional to the quantity $dn/dE$ at $E_F$, i.e., change in carrier concentration with respect to energy at $E_F$, which is expected to vary drastically due to opening of a gap at $E_F$ below the MMT as shown in the previous ARPES studies~\cite{Li2017}.  

We notice that, for $T > $~\tcdw , $|S|$ increase almost linearly with increasing temperature as is typically seen for metals. Naively, one can use the single parabolic band model approximation to rewrite the Mott formula in eq.~\ref{eq1} in the following form:

\begin{equation}
\centering
S = \frac{8\pi ^2 k_B^2 m^*}{3eh^2} \left (  \frac{\pi}{3n} \right ) ^\frac{2}{3} T
\label{eq3}
\end{equation}

where $m^*$ is the band effective mass of the charge carriers. By fitting $S$ above \tcdw\ using $S = a_oT$, where $a_0$ is the prefactor in eq. \ref{eq3}, one can estimate $m^*$. For this purpose, we use $n$ obtained from the Hall coefficient $R_H \approx 10^{-3}$cm$^3$/C at $T = 300$~K \cite{kobayashi1996transport}. Following this procedure, we get $m^* \approx 3.0m_0$ for La$_4$Ni$_3$O$_{10}$,  $\approx 3.7m_0$ for Pr$_4$Ni$_3$O$_{10}$, and $\approx 2.7m_0$ for Nd$_4$Ni$_3$O$_{10}$.
 
\textit{Thermal conductivity}: The temperature variation of thermal conductivity ($\kappa$) is shown in Fig.~\ref{Transport}(d--f). For all three samples the \tcdw\ is manifested in $\kappa$ as a small but clearly discernible kink. For $R =$ La and Pr, we measured the data both while heating and cooling and found some hysteresis around \tcdw. However, since no hysteresis was found in $\rho$, it is difficult to conclude if this is an intrinsic feature or a measurement issue. At low-temperatures, $\kappa$ increases upon heating as $\sim T^3$, which suggests that in this temperature range the acoustic phonons contributes dominantly to $\kappa$. Upon further heating, a noticeable change in the functional form of $\kappa$ takes place for $T \gtrsim 50$~K: In La$_4$Ni$_3$O$_{10}$ $\kappa$ shows a broad peak in the range from $50$~K to $100$~K with a peak value of $3$ Wm$^{-1}$K$^{-1}$ around $80$~K; in Pr$_4$Ni$_3$O$_{10}$ $\kappa$ shows an increasing behavior all the way up to $300$~K, albeit with a much slower rate  $T \gtrsim 50$ K; and, in Nd$_4$Ni$_3$O$_{10}$, $\kappa$ gradually levels off with a saturated value of $\approx1$ Wm$^{-1}$K$^{-1}$ for $T > 100$~K. Thus, the behavior of $\kappa$ in all three cases is rather similar at low-temperatures, but differs somewhat depending on $R$ in the range $T \gtrsim 50$ K. 

It is interesting to note that in spite of their reasonably high electrical conductivities (ranging from $100$-$1000$ S cm$^{-1}$), the thermal conductivities of these nickelates, ranging from $1$ W m$^{-1}$K$^{-1}$ to $3$ Wm$^{-1}$K$^{-1}$, is rather low, which, in turn, implies that the lattice thermal conductivity in these nickelates is intrinsically very low. This may be related to their complex layered structure. The low thermal conductivity and metal-like electrical conductivity above the MMT together indicates that the trilayer nickelates are potential oxide thermoelectric materials.

   \begin{figure*}[t]
	\centering
	\includegraphics[width=2\columnwidth]{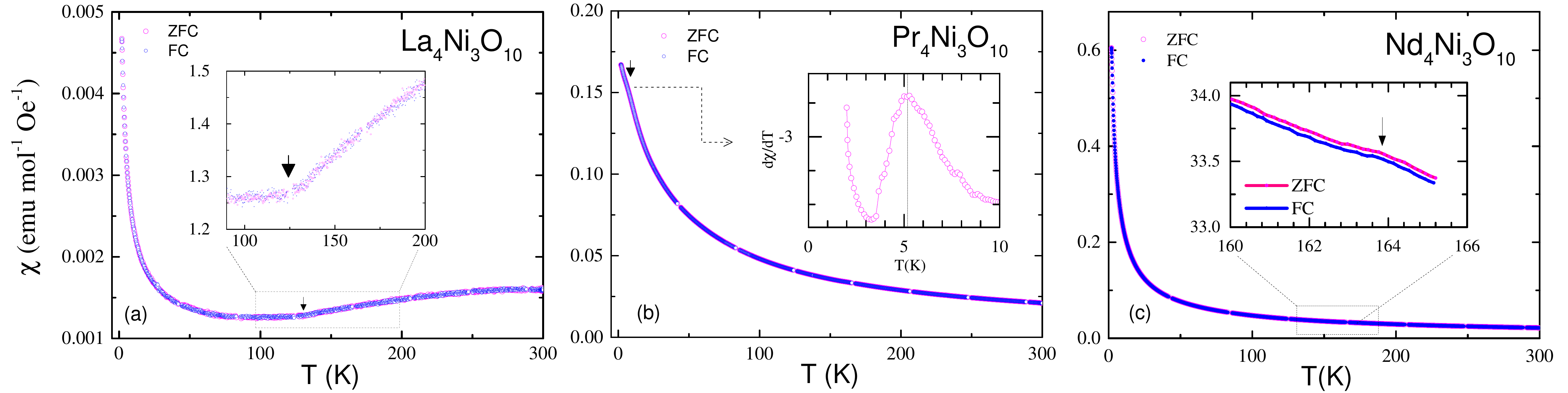}
	\caption{Zero-field-cooled (ZFC) and field-cooled (FC) susceptibility ($\chi$) of (a) \la, (b) \pr, and (c) \nd~measured under an applied field of $5$~kOe. The inset in:  (a) shows a kink in susceptibility at \tcdw, (b) the low-temperature anomaly is emphasized in the first derivative plot, (c) the kink in susceptibility at \tcdw.}
	\label{chi}
\end{figure*}

\textit{Magnetic susceptibility}: The magnetic susceptibility ($\chi$) is shown as Fig. \ref{chi}. Our data shows a good agreement with previous reports \cite{PhysRevB.63.245120, kobayashi1996transport, KUMAR2020165915}. In~\la,~$\chi$ exhibits a discernible kink at $T = 136$~K, which corresponds well with the MMT. In the temperature range $136$ K~$< T < 300$~K, $\chi(T)$ decreases upon cooling which is uncharacteristic of a local moment system. However, at low-temperatures it increases sharply upon cooling. The upturn in $\chi$ could be fitted using the modified Curie--Weiss (CW) law: $\chi = \chi_o + C/(T - \theta_P)$ in the range $2$ K $ \leq $ T $\leq 10$ K yielding $\chi_0 = 10^{-3}$ emu mol$^{-1}$Oe$^{-1}$, the Curie-constant $C = 1.7 \times 10^{-2}$ emu mol$^{-1}$Oe$^{-1}$K$^{-1}$, and the paramagnetic Curie-temperature $\theta_P$ $\approx 2.7$ K. From a previous ARPES study~\cite{Li2017} we know that a gap of $\approx20$ meV opens-up in the $d_{3z^2 - r^2}$ band below~\tcdw, which may induce a localization of $d_{3z^2 - r^2}$ electrons upon cooling leading to the observed upturn below~\tcdw. The overall magnetic behavior of \la~exhibits a complex interplay of itinerant and local moment behavior.
   
The magnetic susceptibility of the~\pr~is dominated by the CW behavior associated with the $Pr^{3+}$ moments. Additionally, a weak anomaly is also observed around $T \simeq 5$~K. The high-temperature $\chi$ could be fitted using the modified CW law yielding: $\chi_0$ $\approx2.8 \times 10^{-3}$ emu mol$^{-1}$Oe$^{-1}$, C $\approx$ $6.3$ emu mol$^{-1}$Oe$^{-1}$K$^{-1}$, and $\theta_p$ $\approx -36$~K in good agreement with literature~\cite{BASSAT1998173}. The value of $\chi_0$ is positive and comparable in magnitude to that for La$_4$Ni$_3$O$_{10}$. The negative sign of $\theta_p$ indicates antiferromagnetic nature of exchange between the Pr$^{3+}$ moments. The experimental effective magnetic moment per formula unit can be estimated using the formula: $\mu_\mathrm{{eff}} = \sqrt{8C}$ which gives $\approx 7.2$ $\mu_B$. Theoretically, $\mu_\mathrm{{eff}}/f.u.$ is given by $[4.\mu^2_{\rm eff}(Pr) + 3.\mu^2_{\rm eff}(Ni)]^\frac{1}{2}$. Substituting the theoretical value of $\mu_\mathrm{{eff}} = 3.58~\mu_B$ per $Pr^{3+}$ ion results in a relatively negligible moment on the Ni-ions.

Though effective magnetic moments of Nd$^{3+}$ and Pr$^{3+}$ are nearly the same in free space, the low-temperature $\chi$ in \nd~is almost four times as large as that of \pr~. This suggests the presence of strong crystal field effect that renders one-half of the Pr-moment effectively non-magnetic at low-temperature due to their singlet ground state as shown later in the manuscript. The CW fit in this case resulted in: $\chi_0$ $\sim$ $3.8\times10^{-3}$ emu mol$^{-1}$Oe$^{-1}$, $C \sim 6.3$ emu mol$^{-1}$Oe$^{-1}$K$^{-1}$, and $\theta_p = -46.5$~K. These values are in close agreement with those recently reported by Li et al.~\cite{PhysRevB.101.195142}. From $C$, the experimental $\mu_\mathrm{{eff}}$ is estimated to be $\approx 7.1\mu_B/f.u.$ which is practically all due to the Nd$^{3+}$, suggesting that the local moment associated with Ni is comparatively negligible. The value of $\theta_p$ is high given the absence of any magnetic ordering, suggesting that a strong magnetic frustration is at play in these nickelates. For more information we refer the reader to Ref. \citenum{SM}.

 \subsection{Specific heat}\label{sectioncp}
\label{SH}
The specific heat ($c_p$) data of $R_4$Ni$_3$O$_{10}$ compounds exhibits a sharp anomaly at their respective MMTs which is particularly pronounced for \pr\ and \nd~, which also show additional anomalies at low temperatures, associated with the rare-earth sublattice. 

\begin{figure}[!]
 	\centering
 	\includegraphics[width = 0.95\columnwidth]{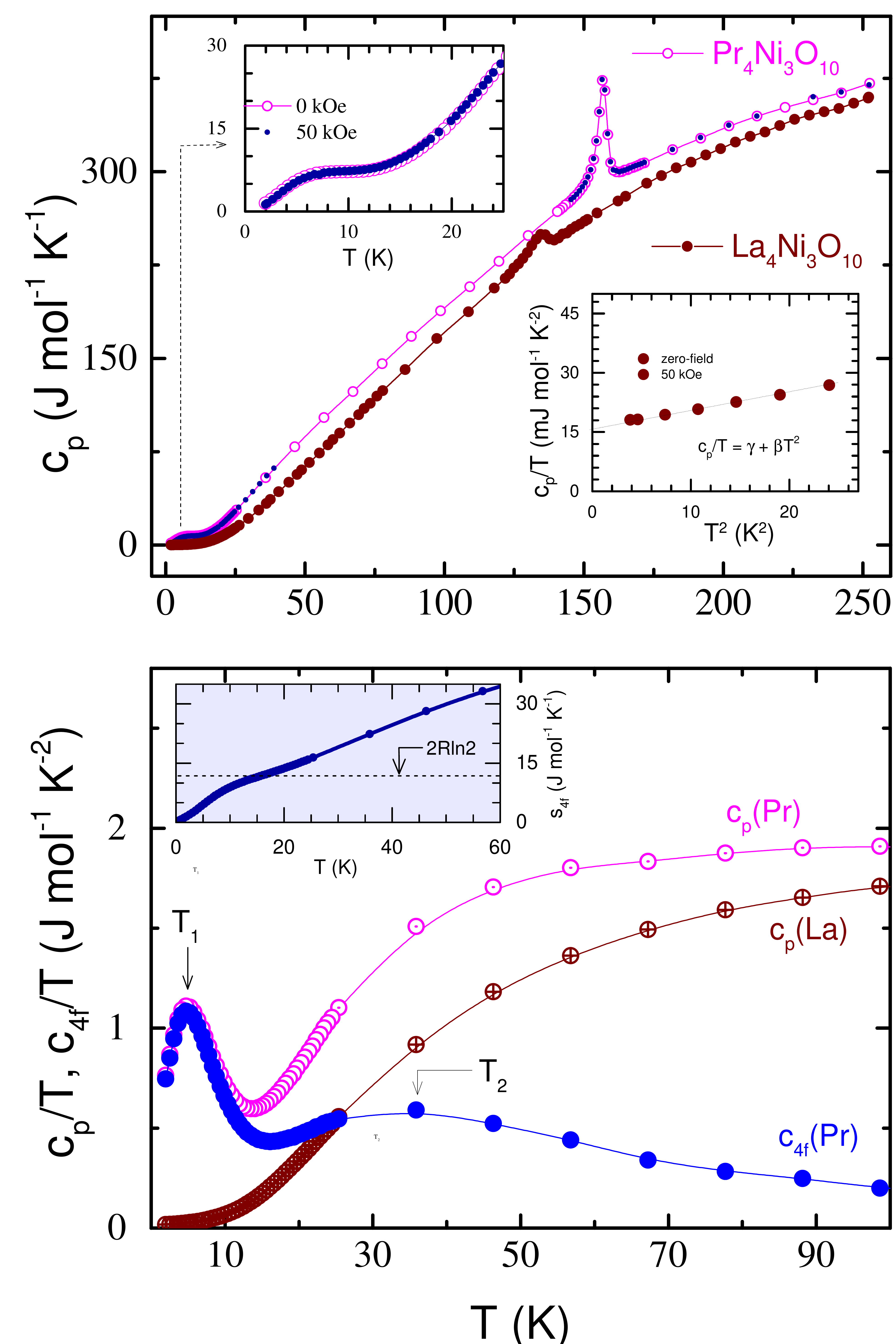}
 	\caption{(a) Temperature ($T$) variation of the specific heat ($c_p$) of \la\ and Pr$_4$Ni$_3$O$_{10}$. The upper inset in (a) highlights the presence of a broad anomaly in $c_p$ of \pr~at low temperature measured under zero-field and  a field of $50$~kOe. The immunity to magnetic field of this peak rules out its Schottky-like origin. The lower inset shows $c_p/T$ vs. $T^2$ of \la\ at low-temperatures. The dashed line is a linear-fit to the data. (b) $\frac{c_p}{T}$ against $T$. $c_{4f}$(Pr) represents the specific heat associated with the $4f$ electrons of Pr$^{3+}$ which is obtained by subtracting the specific heat of \la~ from that of \pr~(see text for details). Temperature variation of entropy  associated with the $4f$ electrons of Pr$^{3+}$ is shown as an inset.}
 	\label{cp_Pr4310}
 \end{figure}

\textit{La$_4$Ni$_3$O$_{10}$}: In La$_4$Ni$_3$O$_{10}$ the specific heat anomaly occurs at $136$~K as shown in Fig.~\ref{cp_Pr4310}. It should be emphasized that in a La$_4$Ni$_3$O$_{10}$ crystallizing in the $Bmab$ space group the specific heat anomaly occurs at a temperature of $\approx150$~K, and it is at $136$ K for the $P2_1/a$ phase \cite{zhang2019high}. This is consistent with our assessment of $P2_1/a$ as the majority phase in our samples. The applied magnetic field of $50$~kOe (not shown) was found to have practically no effect on this anomaly. At low-temperatures, $c_p$ can be fitted using the equation: $c_p= \gamma T + \beta T^3$, where $\gamma$ and $\beta$ represents the electronic and lattice contributions, respectively (see the lower inset in Fig.~\ref{cp_Pr4310}). The best-fit yields: $\gamma \approx 15$~mJ-mol$^{-1}$K$^{-2}$, $\beta \approx0.43$~mJ~mol~K$^{-4}$. The Debye temperature ($\Theta_D$) is calculated from $\beta$ using the relation: $\beta= 12\pi^4Nk_B/5\Theta_D^3$, which gives a value of $\Theta_D \approx 450$~K. The values of $\Theta_D$ and $\gamma$ obtained here are comparable to those previously reported \cite{PhysRevB.63.245120}. From the value of $\gamma$ one can readily estimate the density of states at the Fermi energy, D(E$_F$), using the expression: D(E$_f$)$ = 3\gamma/ \pi^2 k_B^2$, which gives a value of $\approx3.0 \times 10^{22}$ states eV$^{-1}$ cm$^{-3}$. Now, using the carrier density $n$, one can estimate the corresponding density of states D$^{\circ}$(E$_F$) at $E_F$ using the free-electron model. Taking $n \approx6.3 \times 10^{21}$ cm$^{-3}\,$ \cite{kobayashi1996transport}, one gets D$^{\circ}$(E$_F$) $\approx7.6 \times 10^{21}$ states eV$^{-1}$ cm$^{-3}$. From the ratio D(E$_F$)$/$D$^{\circ}$(E$_F$)$= m^*/m_{\circ}$, we estimate the effective mass ($m^*$) for La$_4$Ni$_3$O$_{10}$ to be $m^*\approx3.9 m_{\circ}$, where $m_{\circ}$ is the bare electron mass, which is comparable to the value of $m^*$ from the thermopower ($\approx3.0 m_{\circ}$). The small difference between the two can be due to the possible Fermi surface reconstruction below the MMT. Also, we have not accounted for the valley degeneracy, if any, which makes the the effective mass derived from the density of states higher than the band effective mass by a factor $N^{2/3}$ where $N$ is the valley degeneracy. In any case, the important point is that from the value of $m^*$ one can conclude that the electronic correlations in La$_4$Ni$_3$O$_{10}$ are only modestly enhanced.\\

 \textit{Pr$_4$Ni$_3$O$_{10}$}: Fig.~\ref{cp_Pr4310} shows the specific heat of Pr$_4$Ni$_3$O$_{10}$ where a sharp transition is observed at $156$~K, which agrees nicely with the anomaly associated with the MMT in the transport data. In this case, too, the position and shape of the anomaly remains unaffected by the application of an external magnetic field. Apart from the expected peak at MMT, an additional broad anomaly is seen at low temperatures centered around $T_1 = 5$ K, which coincides with the anomaly in $\chi$ at the same temperature. Interestingly, the applied field up to $50$ kOe has no significant effect on the shape or position of this anomaly ruling out its Schottky-like origin.
\\

To examine the contribution of $4f$ electrons associated with Pr to the specific heat (designated as $c_{4f}^{Pr}$ in the following) at low temperatures, we subtracted the specific heat data of  La$_4$Ni$_3$O$_{10}$ from that of  Pr$_4$Ni$_3$O$_{10}$. Since both are isostructural, with very similar molecular weights, it is, therefore, reasonable to approximate the lattice specific heat of Pr$_4$Ni$_3$O$_{10}$ with that of La$_4$Ni$_3$O$_{10}$. Furthermore, we assume that the small contribution due to Ni 3d electrons to the specific heat does not vary much upon going from La to Pr, at least well below the MMT. This is a reasonable approximation to make given that \tcdw\ of the these nickelates is not very sensitive to the choice of $R$, in wide contrast with the members of the $R$NiO$_3$ series where the structural and physical properties are closed tied to the identity of the rare-earth ion \cite{Catalano_2018}.

 $c_{4f}^{Pr}$ obtained using this procedure is shown in Fig.~\ref{cp_Pr4310}b (lower panel) over the temperature range $2$~K $ \leq T \leq 100$~K. Interestingly, beside the peak at $T_1 = 5$~K, $c_{4f}^{Pr}$ also exhibits an additional broad peak around $T_2 = 36$~K. This new feature is likely a Schottky anomaly arising due to the crystal field splitting of the lowest $J = 4$ multiplet of the Pr$^{3+}$ ions. To understand this further, we estimate the magnetic entropy ($s_{4f}$) buried under the peak at $T_1$ using the formula: $s_{4f} =\int_{0}^{T} \ (c_{4f}^{Pr}/T'\ ) dT'$. For our rough estimate, we extrapolate $c_{4f}^{Pr}$ below $T = 2$~K linearly to $T = 0$~K. The calculated $s_{4f}$ is shown as an inset in the lower panel of Fig. \ref{cp_Pr4310}. It shows a relatively steep rise up to $10$~K, but continues to increase, albeit at a slower rate, upon heating beyond $15$~K. The region between $10$ K and $15$ K is where the crossover from higher ($T < 10$ K) to slower ($T > 15$ K) rates happens. The magnetic entropy released in the temperature range $T \leq 15$~K ($\approx3T_1$) is $\approx 11.5$ J mol$^{-1}$K$^{-1}$, i.e., $\approx2.9$~J Pr-mol$^{-1}$K$^{-1}$, which is approximately $\frac{1}{2}$ of $R\ln2$. What this suggests is that the peak at $T_1$ is likely due to the magnetic ordering of $\frac{1}{2}$ of the Pr$^{3+}$ ions per Pr$_4$Ni$_3$O$_{10}$ formula unit, which is plausible since there are $2$-types of Pr coordinations in this structure: $9$ -- fold (RS layers) and $12$ -- fold (PB layers). Incidentally, Pr$^{3+}$ in the perovskite PrNiO$_3$ has a noon-magnetic singlet ground state~\cite{PhysRevB.60.14857}. Since the coordination of Pr$^{3+}$ ions in the PB layers of Pr$_4$Ni$_3$O$_{10}$ is analogous to that in PrNiO$_3$, it is reasonable to assume that they, too, have a singlet ground state with no magnetic ordering. Therefore, we can tentatively associate the broad peak in the specific heat at  $T_1$ to the magnetic ordering of the $9-$fold coordinated Pr$^{3+}$ ions.  The increase in $s_{4f}$ beyond $15$ K can be attributed to the higher lying crystal field levels as discussed further. A similar scenario has been previously reported for the compounds Pr$_3$RuO$_7$ which has two types of Pr coordinations, namely, eightfold and sevenfold, with Pr ions in the sevenfold coordination having a crystal field split singlet ground state, and those in the eightfold coordination a doublet \cite{PhysRevB.72.014458}. 

 However, the question arises as to why the peak associated with the magnetic ordering of Pr$^{3+}$ ions in the RS layer is not as sharp as is typically seen at a long-range ordered magnetic transitions. To answer this question, one should see that for the $9-$fold coordinated Pr$^{3+}$ ions there are, in fact, two distinct crystallographically sites (Pr$1$ and Pr$2$) as discussed in \ref{IIIA}. Due to minor differences in bond angles and bond lengths around Pr$1$ and Pr$2$, the exchange integrals $J_{11}$ (within the Pr$1$ sublattice), $J_{22}$ (within the Pr$2$ sublattice), and intersite $J_{12}$ may differ slightly, which could be one of the reasons for the $c_p$ anomaly at $T_1$, associated with ordering of Pr$1$ and Pr$2$, to be broad. The other reason could be related to the fact that a $Pr^{3+}$ moments in a RS layer is only weakly coupled to the $Pr^{3+}$ moments in the RS layer above it (see Fig. \ref{CS}), leading to a quasi-two-dimensional behavior.   

Let us now turn our attention to the peak at $T_2$ which seems to arise due to the crystal field splitting of the lowest J-multiplet of Pr$^{3+}$ ions. In a previous inelastic neutron scattering study on the perovskite compound PrNiO$_3$ \cite{PhysRevB.60.14857}, it was found that the $9-$fold degenerate J-multiplet of the $Pr^{3+}$ ion splits into $9$ singlets due to the crystal field effect. The energy difference between the ground state singlet $(E_0^1)$ and the first excited state $(E_1^1)$ is $6.4$ meV or approximately $70$~K. In the first order approximation, the crystal field splitting of Pr ions in the PB layers of Pr$_4$Ni$_3$O$_{10}$ can be assumed to be similar to that in the compound PrNiO$_3$. Within this assumption, the Schottky anomaly due to the ground and first excited singlet is expected to be centered slightly below $T = (E_1^1 - E_0^1)/2k_B \approx 35$~K, which is remarkably close to the position of the peak at $T_2$. Since the second excited singlet for Pr in the PB layers is located around $E_2^1 = 15$ meV ($\approx 165$~K), it is too high up to have any significant effect on the Schottky anomaly arising due to the $E_0^1$/$E_1^1$ pair.\\

It can therefore be concluded that the Pr ions in the PB layer have a singlet ground state due to a crystal field effect, with a broad Schottky anomaly associated with ground and first excited singlet pair. On the other hand, Pr ions in the RS layers have a crystal field split doublet as their ground state, and undergo magnetic ordering around $T_1$. The observed increase in $s_{4f}$ above $2\,T_1$ is partly due to $E_0/E_1$ excitations associated with Pr-ions in the PB layer, and partly due to the higher lying crystal field split levels of Pr ions in the RS layers. In the absence of a detailed crystal field splitting scheme for the Pr ions in the RS layers, a quantitative analysis of the low-temperature specific heat is left as a future exercise.\\

\begin{figure}[!]
  	\centering
  	\includegraphics[width = \columnwidth]{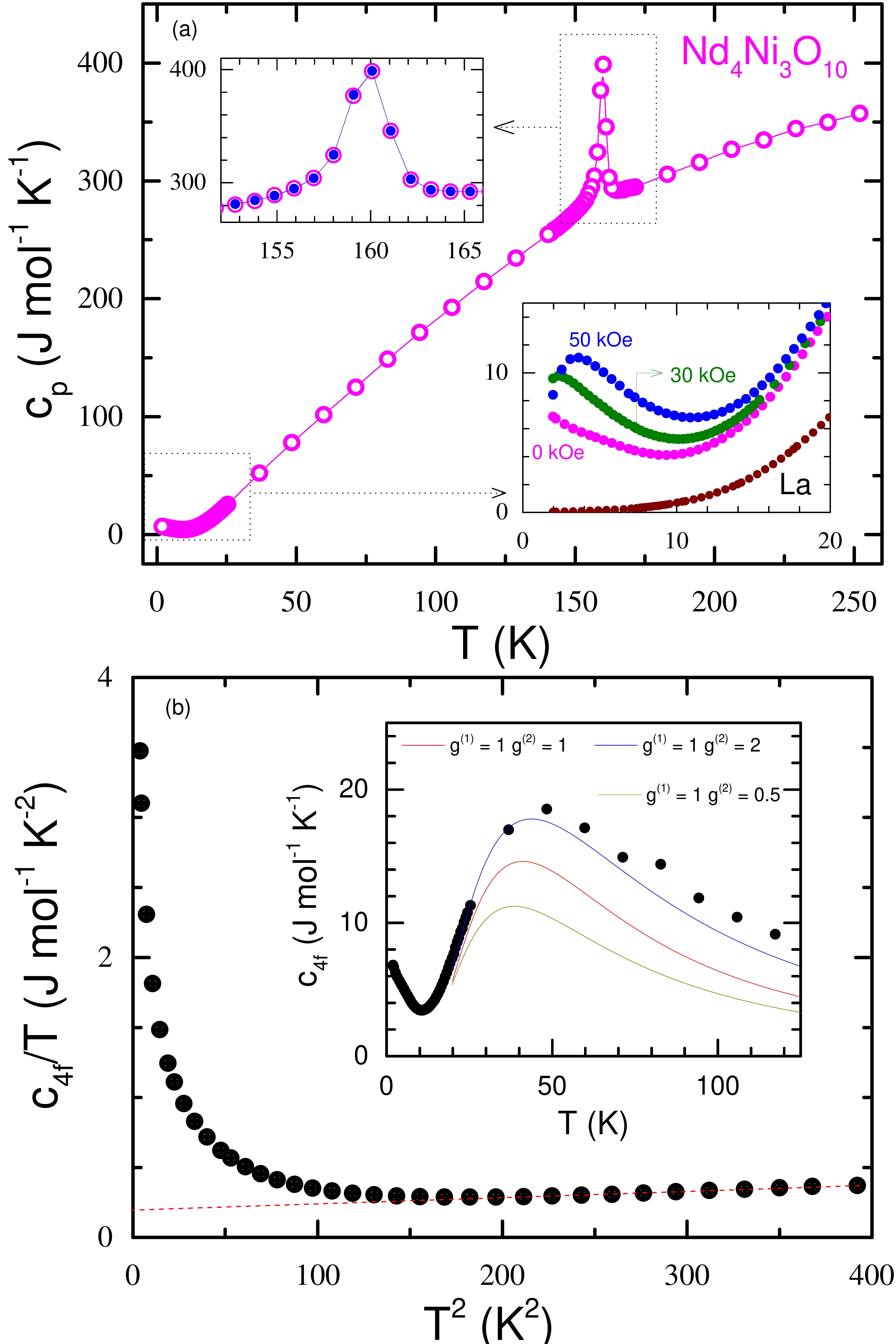}
  	\caption{(a) Specific heat ($c_p$) of Nd$_4$Ni$_3$O$_{10}$. Lower inset shows $c_p$ in the low-temperature range for an applied field of $0$~kOe, $30$~kOe and $50$~kOe; $c_p$ of \la\ is also shown for comparison. Upper inset shows an expanded view of the anomaly at MMT under zero-field and a field of 50 kOe. (b) Low-temperature specific heat associated with the $4f$ electrons of \nd~is plotted as $c_{4f}/T$ versus $T^2$; inset shows $c_{4f}$ versus $T$ up to $T = 120$ K to show the presence of a pronounced Schottky anomaly near $T = 40$ K. The modified Schottky fittings for three cases: $g^{(1)} = 1$, $g^{(2)} = 1$ (red), $g^{(1)} = 1$, $g^{(2)} = 2$ (blue), and $g^{(1)} = 1$, $g^{(2)} = 0.5$ (khaki) (see text for details)}
  	\label{cp_Nd4310}
  \end{figure}

 \textit{\nd}: Fig.~\ref{cp_Nd4310}(a) shows the specific heat of Nd$_4$Ni$_3$O$_{10}$, which is characterized by a sharp anomaly at $T = 160$~K. The position of this anomaly is in a fairly good agreement with the MMT inferred from the transport data, and is found to be independent of an applied magnetic field at least up to $50$~kOe. The low temperature $c_p$ is characterized by an upturn below $T = 10$~K. Under an applied magnetic field, this upturn evolves leading to a broad peak, centered around $4$~K under $H = 50$~kOe, which progressively shifts to higher temperatures with increasing magnetic field. This behavior is reminiscent of a Schottky-like anomaly, which often arises in the rare-earth based compounds due to the crystal field splitting.\\

  To investigate this further, we estimate the specific heat associated with $4f$ electrons of Nd, labeled $c_{4f}^{Nd}$. The specific heat of La$_4$Ni$_3$O$_{10}$ is used as a lattice template, and also to subtract the small magnetic specific heat associated with the Ni sublattice. $c_{4f}^{Nd}$  obtained in this manner is displayed in the lower panel of Fig.~\ref{cp_Nd4310} (inset). At $T = 2$~K, it has a value of about $\sim6.9$ J mol$^{-1}$K$^{-1}$, which decreases sharply upon heating but remains substantial ($\sim3.5$ J mol$^{-1}$K$^{-1}$) even at $T = 12$~K, and increases again upon further heating, exhibiting a broad Schottky like anomaly near $T = 50$~K that can be attributed to the higher-lying crystal field split levels of Nd$^{3+}$ ions. In NdNiO$_3$, for example, the lowest $^4I_{9/2}$ multiplet of Nd$^{3+}$ ion splits into \textit{five} Kramers doublets with the first excited doublet situated around $100$~K above the ground doublet \cite{bartolome1994low}. Since Nd$^{3+}$ ions in the PB layers of Nd$_4$Ni$_3$O$_{10}$ are analogously coordinated, one can assume a similar crystal field splitting scheme for them. On  the other hand, for the $9$--fold coordinated Nd$^{3+}$ ions the splitting scheme may be different. However, since Nd$^{3+}$ is a Kramers ion with $3$ electrons in the $f$--orbitals, in the absence of a magnetic field each crystal field split level should at least be two fold degenerate: i.e., for the $9$--fold coordinated Nd$^{3+}$ ions the ground and first excited state crystal field split levels can have degeneracies as follows:
  $g_0 = 2$, $g_1 = 2$, $g_0 = 2$, $g_1 = 4$, or $g_0 = 4$, $g_1 = 2$. Thus, the ratio $\frac{g_1}{g_0}$, which appears in the expression for the Schottky anomaly, can take values $1$, $2$ or $0.5$, respectively. Note that for $Nd^{3+}$ ions in the PB layer this ratio will be $1$. With this as an input, one can try fitting the broad peak in c$_{4f}$ near $40$~K using the expression: $c_{Sch} = c_{Sch}^{(1)} + c_{Sch}^{(2)}$, where:

  \begin{equation}
  \centering
  \
  c_{Sch}^{(i)} = 2R\left( \frac{\Delta_i}{T} \right)^2 \frac{g^{(i)}exp(\frac{-\Delta_i}{T})}{[1 + g^{(i)}exp(\frac{-\Delta_i}{T})]^2}
  \
  \label{eq5}
  \end{equation}

  In this expression, $R$ is the universal gas constant, $\Delta$ is the splitting between the ground and first excited state, and $g$  is the ratio $\frac{g_1}{g_0}$. Here, the index $i$ is used for the the two types of coordinations, \textit{viz}, $i = 1$ corresponding to the $12$--fold coordination, and $i = 2$ corresponding to the $9$--fold. The prefactor $2$ account for the number of Nd$^{3+}$ ions per formula unit in each type of layers. The fitting result for $g^{(1)} = 1$, $g^{(2)} = 1$ (fit$1$), $g^{(1)} = 1$, $g^{(2)} = 2$ (fit$2$), and $g^{(1)} = 1$, $g^{(2)} = 0.5$ (fit$3$) are shown in the inset of Fig.~\ref{cp_Nd4310}(b). The corresponding values of $\Delta_1$ and $\Delta_2$ for these fits are: $98$ K and $98$ K for (fit$1$), $150$ K and $95$ K for (fit$2$), and $87$ K and $93$ K for (fit$3$), respectively. Clearly, the best fit corresponds to (fit$2$), which implies that the ground state of Nd$^{3+}$ ions in the $9$--fold coordination is also a Kramers doublet, with a quartet for the first excited state.\\

  Let us now turn our attention to the increase in  $c_{4f}^{Nd}$ upon cooling below $T = 10$~K. In NdNiO$_3$ a similar upturn leading to a broad peak around $T = 1.7$~K had been previously reported \cite{bartolome1994low}. It was argued to arise from the exchange splitting of the ground state doublet. However, unlike NdNiO$_3$, in Nd$_4$Ni$_3$O$_{10}$ the Ni moments are not ordered and hence the Ni--Nd exchange field in this case is almost non-existent. On the other hand, it might be that this upturn is precursory to an impending magnetic ordering of the Nd moments at further low-temperatures. After all, the Nd-Nd exchange, as inferred from the high temperature Curie-Weiss fit, is about $-45$~K, which is rather high. This could then be a case closely analogous to the case of Nd$_2$O$_3$ recently reported, which also exhibits a high $\theta_p \simeq -24$~K, but with long-range order setting in only below $T = 0.55$~K. Surprisingly, $c_p$ of Nd$_2$O$_3$ shows not only a sharp peak at $0.55$ K corresponding to the long-range ordering of Nd moments but also a broad feature centered around $1.5$~K. The authors report that the entropy associated with this broad peak must be taken into account in order to recover the $R\ln2$ entropy expected from a ground state doublet suggesting a complex two-step ordering of the Nd moments. The $c_p$ of Nd$_4$Ni$_3$O$_{10}$ also shows a broad peak at $T \approx 1.8$ K \cite{PhysRevB.101.195142} which suggests that a phenomenology analogous to Nd$_2$O$_3$ might also be at play here. Further studies down to much lower temperatures would be interesting to explore this analogy further and to understand the true ground state of the Nd sublattice. \\

  Finally, $c_{4f}^{Nd}/T$ versus $T^2$ is plotted in the lower panel of Fig.~\ref{cp_Nd4310}. The data from $12$~K to $20$~K can be fitted to a straight line whose intercept on the $y-$axis is $\sim150$ mJ mol$^{-1}$K$^{-2}$. Indeed, in Ref. \citenum{PhysRevB.101.195142}, a high $\gamma$ value of $146$ mJ mol$^{-1}$K$^{-2}$ is reported by fitting $c_p/T$ versus $T^2$ to $\gamma + \beta T^2$ in this temperature range. However, caution must be exercised while interpreting the intercept value in this case since $c_p$ in this temperature range, as shown in the inset of Fig. \ref{cp_Nd4310}b, is overwhelmed by the Schottky contribution arising from the crystal field split lowest $J-$multiplet of Nd$^{3+}$ ions. It is for this reason we believe that the erroneously high $\gamma$ value in Ref.~\citenum{PhysRevB.101.195142} misled the authors to conclude a "novel" heavy-electron behavior in Nd$_4$Ni$_3$O$_{10}$, is a gross overestimation. In fact, as shown in the supplementary~\cite{SM}, if one use the same procedure for~\pr, a high $\gamma$ value of $\approx300$ mJ mol$^{-1}$K$^{-2}$ will emerge, but we know from the work of Huangfu et al.\cite{HuangfuPRR2020}, the resistivity of a~\pr single crystal decreases upon cooling at low-temperature, i.e., no heavy-electron behavior is observed in the transport studies. However, as is well documented in the heavy fermion-literature, if the electronic specific heat $\gamma$ in such cases is derived by extrapolating the high-temperature specific heat data to $T=0$~K using $\gamma T + \beta T^2$ unusually large values emerge, which can be \textit{falsely} interpreted as arising due to the heavy fermion behavior.

 \subsection{Thermal Expansion and Gr\"{u}neisen analysis}
 \label{TE}
\begin{figure}[htb]
\centering
\includegraphics [width=0.95\columnwidth,clip] {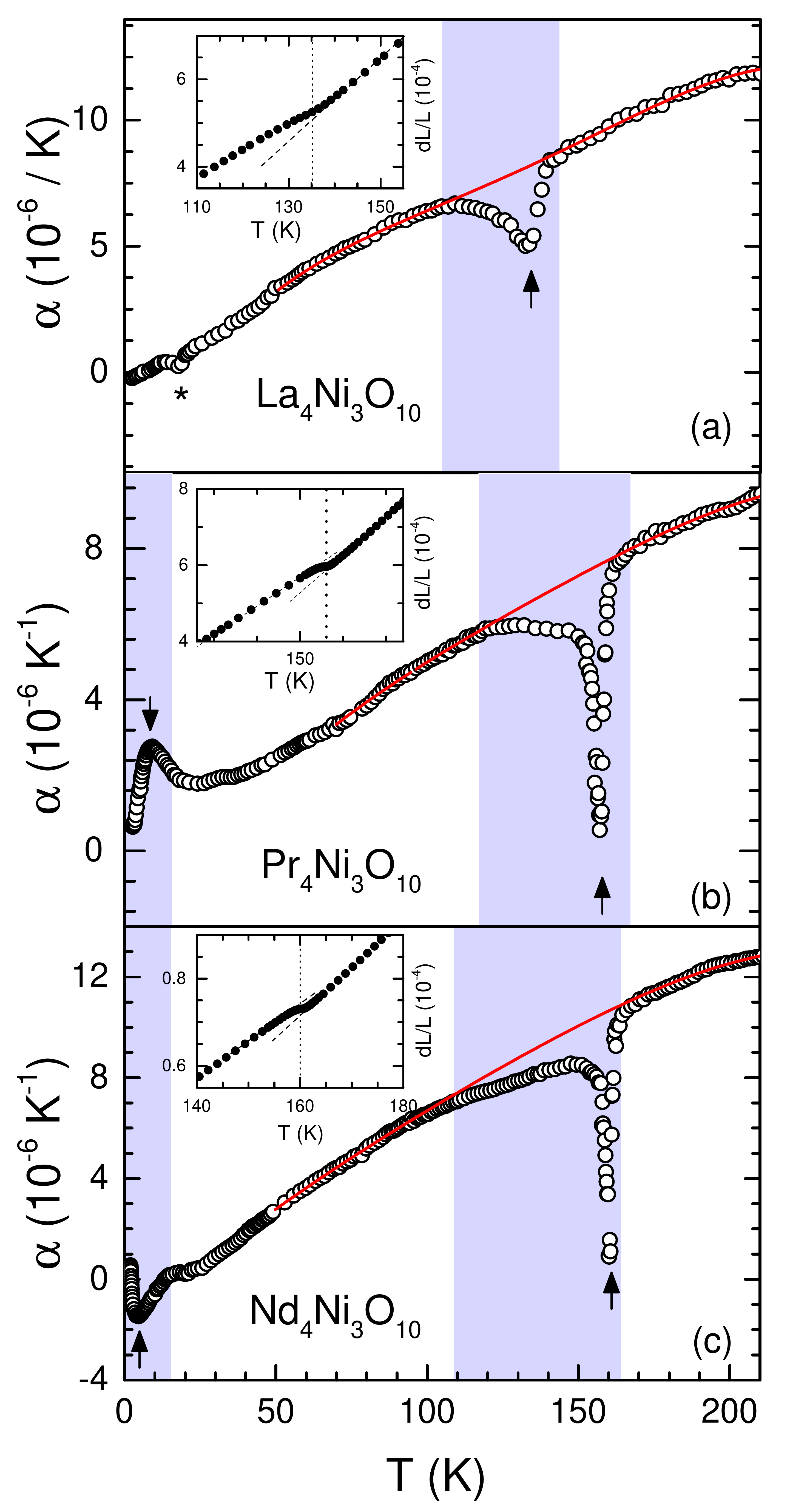}
\caption{Temperature dependence of the thermal expansion coefficient $\alpha$ of \la, \pr, and \nd. The red line shows a polynomial estimate of the background (see text for details). The arrows marks the position of \tcdw. The asterisk in the upper panel indicates an experimental artifact. The additional low-temperature peak in \pr\ and \nd\ is likely due to the crystal field excitations.The inset shows length change ($dL/L$) around \tcdw; the dotted line is a guide to the eye.} \label{all-TE}
\end{figure}

The temperature dependence of length changes studied by capacitive dilatometry, shown in Figs.~\ref{all-TE}, follows the volume dependence as measured using the X-ray diffraction data (see Fig.~\ref{RR4310}(f1, f2 and f3)). However, while there is quantitative agreement for \pr, discrepancies are noticed for \la\ and \nd. Specifically, the dilatometric length changes are about $25$\% and $45$\% larger than suggested by X-ray diffraction, respectively. The data are isotropic, i.e., we find the behavior to be the same when measuring along different directions of the polycrystalline cuboids, which excludes a simple non-random orientation effect to cause this discrepancy. Instead, the data suggest a non-uniform internal stress distribution within the polycrystalline samples which can lead, in porous materials, to larger thermal expansion than in the bulk~\cite{ho1998thermal}.\\

The length changes in \re\ evidence significant coupling of electronic and structural degrees of freedom. Specifically, there are pronounced anomalies at \tcdw\ in all studied materials. In \la , the data in Fig.~\ref{all-TE} displays a broad feature which signals shrinking of the sample volume upon exiting the MMT phase while heating the sample. Qualitatively, this implies negative hydrostatic pressure dependence $dT_{\rm MMT}/dp < 0$. The minimum of the thermal expansion anomaly appears at \tcdw\ $ = 134$~K, suggesting either a weak first-order character of the transition or a somehow truncated $\lambda$-like behavior similar to what is indicated by the specific heat anomaly (cf. Fig.~\ref{cp_Pr4310}).

\begin{figure}[htb]
\centering
\includegraphics [width=0.95\columnwidth,clip] {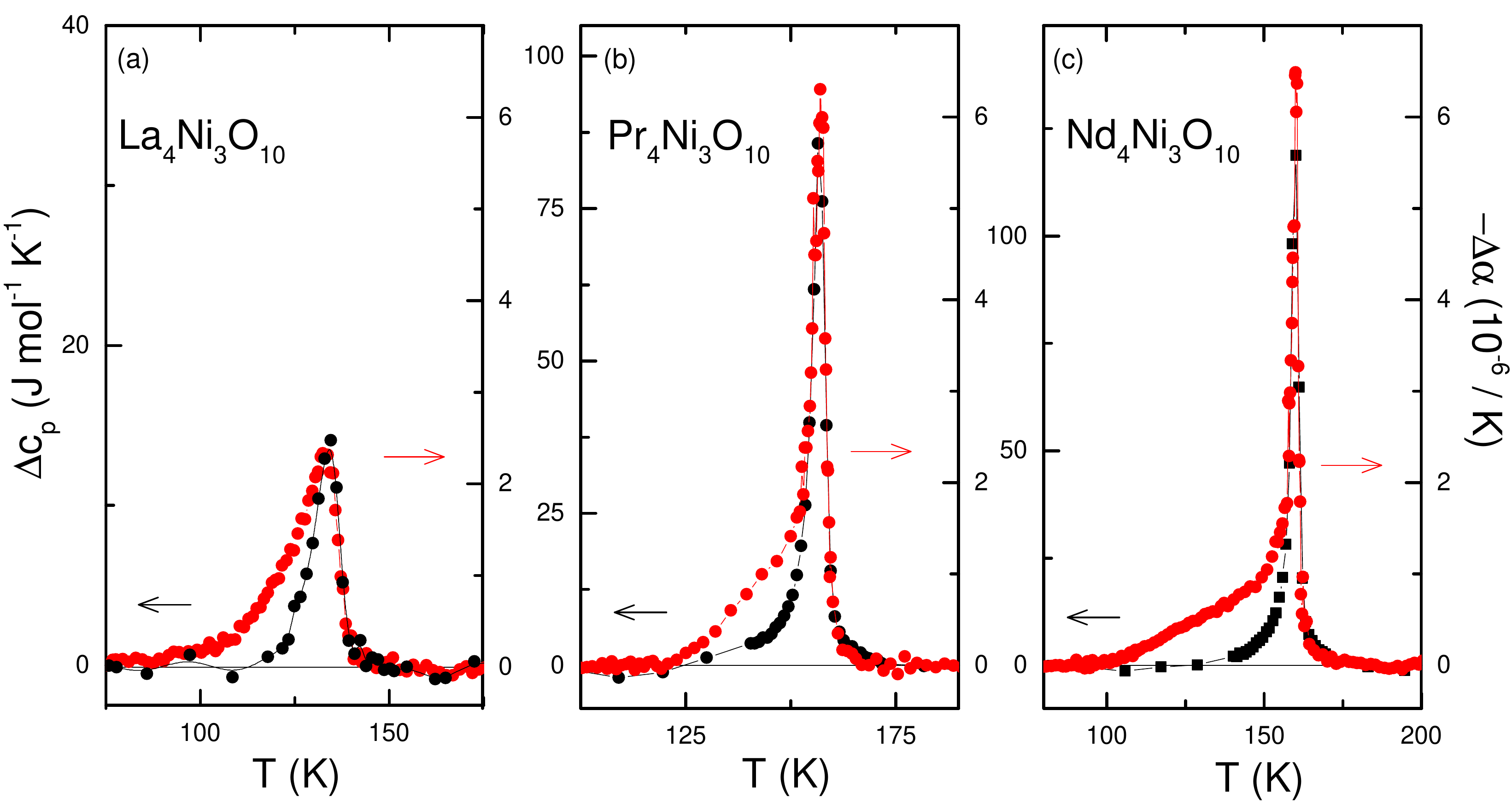}
\caption{Anomalies in the specific heat and the negative thermal expansion coefficient of \re\ with $R$ = La, Pr, and Nd. The anomaly size in (a) and (c) has been rescaled according to the X-ray diffraction results (see the text). Note the same scale of the thermal expansion ordinate in all graphs. } \label{grueneisen}
\end{figure}

In order to estimate the background contribution to the thermal expansion coefficient, a polynomial was fitted to the data well below and above the thermal expansion anomaly as shown in Fig.~\ref{all-TE}~\cite{PhysRevB.65.174404}. The background \alb\ mainly reflects the phonon contribution. Due to the large size of the anomaly, using  different temperature ranges for the determination of the background and/or choosing  different  fit  functions  does  not  change  the result significantly. Subtracting \alb\ from the data yields the anomaly contribution to the thermal expansion coefficient $\Delta \alpha$ as shown in Fig.~\ref{grueneisen}a. Recalling the discrepancy of dilatometric and XRD length changes mentioned above for \la\ and \nd , for the following quantitative analysis of both we have scaled the dilatometric data to the XRD results. Quantitatively, our analysis then yields total anomalous length changes $\Delta_t L/L = \int \Delta \alpha dT = -4.2(9) \cdot 10^{-5}$.

\begin{table*}[!]
	\setlength{\tabcolsep}{4pt}
	\caption{Total anomalous length and entropy changes $\Delta_t L/L = \int \Delta \alpha dT$ and $\Delta_t S = \int \Delta c_{\rm p}^{\rm MMT}/T dT$, discontinuous length changes $\Delta_d L/L$, Gr\"{u}neisen parameter $\Gamma$ and hydrostatic pressure dependence of \tcdw\ of \re\ (see the text).}
\label{tab1}
\begin{center}
	\begin{tabular}{c c c c c c}
		
		\hline\hline
		&  $\Delta_t L/L$ &  $\Delta_d L/L$ &  $\Delta_t S$ & $\Gamma$ & $dT_{\rm MMT}/dp$ \\
		\hline
		\tabularnewline
		\la\ & $-4(1)\cdot 10^{-5}$ & - & 1.0(3)~\jmk  & $-4.9(9)\cdot 10^{-7}$~mol/J & $-8(2)$~K/GPa \\
		\pr\ & $-5(1)\cdot 10^{-5}$   & $-3.1(6)\cdot 10^{-5}$& 3.1(6)~\jmk  & $-2.3(6)\cdot 10^{-7}$~mol/J & $-4(1)$~K/GPa \\
		\nd\ & $-5.1(4)\cdot 10^{-5}$ & $-2.6(2)\cdot 10^{-5}$ & 3.5(9)~\jmk & $-1.4(4)\cdot 10^{-7}$~mol/J & $-3(1)$~K/GPa \\
		\tabularnewline
		\hline\hline
		
	\end{tabular}	
\end{center}
\end{table*}

When replacing La by Pr and Nd in \re , the anomalies in the thermal expansion at \tcdw\ become significantly sharper and evidence rather discontinuous behavior (see Figs.~\ref{all-TE}). In addition, there are pronounced features at low temperatures (marked by arrows) that are associated with the rare-earth sublattice. In particular, the data clearly confirm negative volume expansion in \nd\ below $\sim 20$~K. At higher temperatures, the sharp anomalies at $156$~K ($R$ = Pr) and 160~K ($R$ = Nd) at \tcdw\ are accompanied by a regime of rather continuous length changes which extends from \tcdw\ down to about $110$ K, i.e., it is significantly larger than the anomaly regime in \la . Applying the procedure described above for determining the background yields the thermal expansion anomalies as displayed in Fig.~\ref{grueneisen}b and \ref{grueneisen}c for the two compounds.

The anomalies $\Delta \alpha$ in the thermal expansion coefficients at \tcdw \: are presented in Fig.~\ref{grueneisen} together with the respective anomalies of the specific heat. The latter have been derived by estimating the background specific heat analogously to the procedure used for the thermal expansion data and by using the same fitting regimes in both cases~\cite{PhysRevB.65.174404}. For each composition, scaling of \dcp\ and $\Delta \alpha$ has been chosen to obtain the best overlap of the specific heat and thermal expansion data around \tcdw\ and above. The fact that the thermal expansion and specific heat anomalies are proportional at \tcdw\ implies a $T$-independent Gr\"{u}neisen parameter describing the ratio of pressure and temperature dependence of entropy changes in this temperature range. This observation implies the presence of a single dominant energy scale $\epsilon$~\cite{Gegenwart_2016}. In contrast, the fact that Gr\"{u}neisen scaling starts to fail at around $10$~K below \tcdw\ indicates the presence of more than one relevant degree of freedom. In the temperature regime around \tcdw\ and above, the corresponding scaling parameter is the Gr\"{u}neisen parameter~\cite{PhysRevB.73.214432}:

$$\Gamma = \frac{3\Delta \alpha}{\Delta c_{\rm p}} = \frac{1}{V}\left. \frac{\partial \ln \epsilon}{dp}\right|_T .$$

 Our analysis yields the $\Gamma$ values summarized in Table~\ref{tab1}. Using the Ehrenfest relation, the obtained values of $\Gamma$ yield the hydrostatic pressure dependencies of the ordering temperature at vanishing pressure, i.e., $dT_{\rm MMT}/dp = T_{\rm MMT} V_{\rm m}\Gamma$. The results deduced using the molar volume $V_{\rm m}$ are shown in Table~\ref{tab1}.

The obtained initial slopes of hydrostatic pressure dependencies of \tcdw\ are comparable to values reported from measurements of the electrical resistivity under pressure. Specifically, Wu~\etal\ report $-6.9$~K/GPa for \la\ which nicely agrees to the results of the Gr\"{u}neisen analysis presented above. The comparison with \nd\ studied in Ref.~\citenum{PhysRevB.101.195142} is, however, ambiguous. On the one hand, Li et al. ~\cite{PhysRevB.101.195142} report discontinuous shrinking of the unit cell volume at \tcdw\ by $0.08$~\% while cooling, which, both, qualitatively and quantitatively, contrasts our data (cf.~inset of Fig.~\ref{all-TE}c). In particular, this value implies a $positive$ hydrostatic pressure dependence of about $+35$~K/GPa~\footnote{We have applied the Clausius-Clapeyron equation and used $\Delta S=2.8$~\jmk\ as reported in Ref.~\cite{PhysRevB.101.195142}.}. However, at the same time an initial $negative$ hydrostatic pressure dependence of about $-8$~K/GPa is reported in Ref.~\citenum{PhysRevB.101.195142} which thermodynamically contradicts the reported volume changes at \tcdw\ but is reasonably consistent with the results of our Gr\"{u}neisen analysis.

The broad region of anomalous length changes between \tcdw\ and $\sim100$~K signals clear temperature variation of the Gr\"{u}neisen ratio, in this temperature regime, the reason of which is not fully clear. In general, the fact that capacitance dilatometry is obtained under small but finite pressure, which in the case at hand is estimated to about $0.6(1)$~MPa, may affect measurements in particular on polycrystalline samples. The fact that the dilatometer detects volume increase however renders a scenario as observed in recent studies of electronic nematicity of LaFeAsO rather unlikely, where the shear modulus $C_{66}$ is the elastic soft mode of the associated nematic transition so that dilatometry under finite pressure results in associated volume decrease~\cite{PhysRevB.80.094512,PhysRevLett.105.157003}. We also exclude that variation of $\Gamma$ is associated with incompletely resolved strain from the discontinuous transition at \tcdw\ because the measurements have been performed upon heating and the temperature regime of the observed anomaly is very large. Instead, we conclude the presence of a competing ordering phenomenon as suggested by the failure of Gr\"{u}neisen scaling~\cite{Gegenwart_2016}. Intriguingly, a temperature regime of unexpected behavior has also been detected in the out-of-plane resistivity $\rho_{\perp}$ in \pr\ single crystal where, in contrast to the in-plane resistivity, an increase of $\rho_{\perp}$ upon cooling, i.e., insulating behavior, is observed in a large temperature regime \cite{PhysRevB.101.104104}. It is tempting to trace back this intermediate temperature regime of $d\rho_{\perp}/dT<0$, i.e., a metal-to-insulator-like behavior of $\rho_{\perp}$ at \tcdw , to the competing degree of freedom which manifests in the thermal expansion coefficient and change of Gr\"{u}neisen parameter shown in Fig.~\ref{grueneisen}b.

\section{Summary \& Conclusions}
\label{SC}

We investigated the trilayer nickelates \re~($R= $ La, Pr and Nd) that are $n = 3$ members of the RP series. We focused our investigations on understanding the following important aspects concerning the physical properties of these compounds: (i) what is the correct space group characterizing the room-temperature crystal structure of these compounds, (ii) is there a structural phase transition at \tcdw, (iii)  how do various thermodynamic quantities, including resistivity, magnetic susceptibility, specific heat, thermopower, thermal conductivity and thermal expansion coefficient vary across MMT, and (iv) to understand the magnetic behavior of the rare-earth sublattices in \pr\ and \nd .

In order to address these questions, we synthesized high-quality samples using the sol-gel method. These samples were then subject to a high-resolution synchrotron powder X-ray diffraction at the ALBA synchrotron source, both at 300 K and lower temperatures down to $90$ K. A thorough analysis confirms that these compounds crystallize in the monoclinic $P2_1/a, Z= 4$ phase. Absence of new peaks emerging or splitting of the existing peaks ruled out any lowering of the lattice symmetry accompanying this transition. The thermal expansion coefficient also captured the anomaly at \tcdw\ rather vividly. From the analysis of $\Delta \alpha$, we conclude that the MMT anomaly becomes more first order-like as we go to smaller lanthanide ionic radii (and thereby larger distortions from the perovskite structure).This was further corroborated by temperature variation of various physical properties. 

Resistivity data of all samples exhibit sharp jump or discontinuity at their respective \tcdw\, and an upturn, i.e., with $d\rho/dT < 0$, at low-temperatures. We show that this upturn is likely a consequence of weak-localization arising due to inelastic electron-electron interactions. This result is in agreement with Ref.~\onlinecite{KUMAR2020165915} where resistivity of \la\ has been analyzed in considerable details. In particular, we excluded a Kondo-like mechanism in the Ni-sublattice leading to $d\rho/dT < 0$ as has been proposed recently \cite{PhysRevB.101.195142}. This result is further strengthened by thermopower and specific heat experiments. From thermopower and specific heat, we found the band effective mass of the charge carriers to range from around $3 m_\circ$, which indicates that the electronic correlations are at best  moderately enhanced. 

The magnetic ground state of the $R-$ions in \pr\ and \nd is shown to be rather interesting. First, the Curie-Weiss temperature ($\theta_p$) for both these compounds is of the order of $-40$ K; however the long-range ordering remains suppressed down to temperatures as low as $5$ K for \pr\, and less than $2$ K for \nd\, suggesting the presence of strong magnetic frustration, which may be related to their layered structure that renders the $R^{3+}$ moments located in the RS layers quasi-two-dimensional. From the analysis of $c_p$ and $\chi$, we infer that in \pr, the Pr$^{3+}$ ions located in the PB layers exhibit a crystal field split non-magnetic singlet ground state, while those located in the RS layers show a ground state doublet with an antiferromagnetic ordering below about $5$ K. 

In \nd, on the other hand, all four Nd-ions in the formula unit exhibit a Kramers doublet ground state with first excited state as doublet for one-half of the Nd ions and quartet for the remaining half, giving rise to a pronounced Schottky-type anomaly centered around $T = 35$ K. The low-temperature specific heat of both \pr\ and \nd\ is found to be overwhelmed by the Schottky-like contributions arising from the crystal field excitations associated with the lowest $J$--multiplet of the rare-earth ions, which tends to \textit{falsely} inflate the value of $\gamma$. 

In summary, the rare-earth sublattice in \re\ compounds with $R=$ Pr and Nd, exhibit very intriguing behavior which should be subject to further examination using specific heat down to much lower temperatures, inelastic and elastic neutron scattering. With the possibility of single-crystal growth, the interesting low-temperature behavior of these compounds as shown here should attract significant further interest.\\

\section*{Acknowledgments}
The powder x-ray diffraction experiments were performed at MSPD - BLO4 beamline at ALBA Synchrotron with the collaboration of ALBA staff. The authors thank the Department of Science and Technology, India (SR/NM/Z-07/2015) for the financial support and Jawaharlal Nehru Centre for Advanced Scientific Research (JNCASR) for managing the project. SS acknowledges financial support form SERB (WMR/2016/003792). RK acknowledges support by Deutsche  Forschungsgemeinschaft  (DFG)  through  KL  1824/13-1 and by BMBF via SpinFun (no. 13XP5088).

\bibliography{Rout_etal}

  \end{document}